\documentclass[12pt]{article}
\usepackage{epsf}
\usepackage{graphicx}
\newcommand{\be}{\begin{equation}}
\newcommand{\ee}{\end{equation}}
\newcommand{\bea}{\begin{eqnarray}}
\newcommand{\eea}{\end{eqnarray}}
\newcommand{\ba}{\begin{array}{ccc}}
\newcommand{\ea}{\end{array}}
\newcommand{\nn}{\nonumber}
\newcommand{\noi}{\vspace{12pt}\noindent}

\newcommand{\eq}[1]{eq.~(\ref{#1})}
\newcommand{\eqs}[2]{eqs.(\ref{#1}, \ref{#2})}

\newcommand{\ur}[1]{(\ref{#1})}
\newcommand{\urs}[2]{(\ref{#1},\ref{#2})}

\def\log{\textnormal{log}}
\def\det{\textnormal{det}}
\def\exp{\textnormal{exp}}

\def\gtwid{\raise.3ex\hbox{$>$\kern-.75em\lower1ex\hbox{$\sim$}}}
\def\ltwid{\raise.3ex\hbox{$<$\kern-.75em\lower1ex\hbox{$\sim$}}}
\def\ok{\omega_k}
\def\ge{\gamma_{{\rm E}}}
\def\po{\psi}
\def\Tr{ {\rm Tr} }
\def\Sp{{\rm Sp}}

\def\ch{{\rm ch}}
\def\cos{{\rm cos}}

\def\sin{{\rm sin}}
\def\A{\mathcal{A}}

\begin{document}

\topmargin -1.4cm
\oddsidemargin -0.8cm
\evensidemargin -0.8cm
\title{\Large{{\bf Covariant derivative expansion of Yang-Mills effective action at high
temperatures}
\\ 
}}

\vspace{1.5cm}

\author{~\\{\sc Dmitri Diakonov} \\ 
{\small NORDITA,  Blegdamsvej 17, DK-2100 Copenhagen,
Denmark\footnote{E-mail: diakonov@nordita.dk}}\\ \\
{\sc Michaela Oswald}\\
{\small NBI, Blegdamsvej 17, 2100 Copenhagen, Denmark}\\
{\small and Department of Physics, Brookhaven National Laboratory, Upton, NY 11973-5000, USA\footnote{E-mail: oswald@alf.nbi.dk}}
}

\date{} 
\maketitle
\vfill
\begin{abstract} 
\noindent
Integrating out fast varying quantum fluctuations about Yang--Mills fields $A_i$ and $A_4$,
we arrive at the effective action for those fields at high temperatures. Assuming that the fields
$A_i$ and $A_4$ are slowly varying but that the amplitude of $A_4$ is arbitrary, we find a
nontrivial effective gauge invariant action both in the electric and magnetic sectors. Our
results can be used for studying correlation functions at high temperatures beyond the
dimensional reduction approximation, as well as for estimating quantum weights of classical
static configurations such as dyons.
\end{abstract}


\vfill

\thispagestyle{empty}
\newpage

\section{Introduction}

The range of medium temperatures is probably one of the most interesting aspects
of quantum chromodynamics (QCD). It is the region where the
confinement-deconfinement phase transition is expected in the pure-glue or quenched
variants of the theory, and where chiral symmetry restoration is believed to occur in the
full version, with light dynamical fermions. 
Pure-glue theories without dynamical quarks have the advantage that one can characterize the order parameter and get insight into many interesting aspects of the phase transition \cite{P, S, 'tH, SY}.
To get a good theoretical understanding of what is
going on below and above the phase transitions and to understand the microscopic
mechanism of the transitions themselves, is one of the greatest challenges in QCD. 

Unfortunately, the present theoretical tools to handle these problems are insufficient:
there are a precious few well-based statements about high and intermediate temperatures. 
At very high temperatures the perturbation theory in the running coupling constant
can be developed. Especially the hard-thermal-loop resummation method \cite{HTL} proved essential. However, perturbation theory necessarily explodes already in 
a few-loop approximation due to the nonperturbative chromomagnetic sector of non-Abelian gauge theories \cite{P,L,GPY}, thus limiting the applicability of perturbation theory to
academically high temperatures \cite{breakdown}. The 1-loop \cite{GPY,Weiss} and 2-loop \cite{Belyaev}
potential energies as functions of the `time' Yang--Mills component $A_4$ are known, which are periodic functions with a period $2\pi T$ of the eigenvalues of $A_4$ in the adjoint
representation. The curvature of this potential gives the Debye mass. The potential has
zero-energy minima for quantized values of $A_4$ corresponding to the Polyakov line
assuming values from the center of the gauge group. At high temperatures the system
oscillates around one of those trivial values of the Polyakov line. 

At lower temperatures the fluctuations in the values of the Polyakov line increase and
eventually the system undergoes a transition to the phase with $\Tr\,P=0$, known as the
confinement phase. To study this phase transition or at least to approach it from the
high-temperature side, one needs to know the effective action for the Polyakov line in
the whole range of its possible variation. Effective Lagrangians for $A_4$ at high temperatures have been constructed and studied by a number of authors \cite{coupling, effact}, however, the 1-loop kinetic energy for the Polyakov line is unknown. One of the aims of this paper is to find it.

Let us formulate the problem more mathematically. Nonzero temperatures explicitly break
the $4D$ Euclidean symmetry of the theory down to the $3D$ Euclidean symmetry, 
so that the spatial $A_i$ and time $A_4$ components of the Yang--Mills field play different
roles and should be treated differently. One can always choose a gauge where $A_4$ is
time-independent. Taking $A_4(x)$ to be static is not a restriction of any kind on the fields
but merely a convenient gauge choice, and we shall imply this gauge throughout the paper. 
[It is also a possible gauge choice at $T=0$ but in that limiting case it is unnatural as one
usually wishes to preserve the $4D$ symmetry.] As to the spatial components $A_i(x,t)$, they
are, generally speaking, timedependent, although periodic in the time direction. 
Putting the components $A_i$ to zero is a gauge noninvariant restriction on the fields
since any time-independent gauge transformation will generate a nonzero $A_i$. Therefore,
the spatial derivatives of the Polyakov line in the gauge-invariant effective action can only
appear as {\em covariant derivatives} including a nonzero $A_i$ field. 

The effective action studied in this paper is a functional of the background static $A_4$
field and, generally speaking, nonstatic $A_i$ fields, obtained by integrating out fast-varying
quantum oscillations about the background. The key ingredient is that we do not assume
$A_4$ to be small but sum up all powers in $A_4$. Therefore we are actually computing the
effective action for the Polyakov loop interacting, in a covariant way, with the spatial $A_i$
fields. The resulting effective action has to be invariant with respect to time-independent
gauge transformations and also with respect to certain residual timedependent gauge
transformations which do not induce nonstatic $A_4$ and support the periodicity of 
$A_i(x,t)$; they will be discussed at the end of the paper.  

An economic and \ae sthetic method of getting explicitly gauge invariant actions is based
on the evaluation of functional determinants. [An equivalent method is computing 1-loop
Feynman graphs with arbitrary number of external legs, however it is technically more
involved and does not automatically support gauge invariance with respect to the external
field.] In this case, the evaluation of functional determinants is nontrivial as we expand
it in the (covariant) derivatives of the field but sum up all powers of the amplitude of $A_4$.
We develop a general technique for the covariant derivative expansion which, in principle,
can be worked out to any power of the derivatives. In this paper, however, we find
explicit expressions for the action with 0,2 and 4 covariant derivatives. This enables us
to find the leading terms both in the electric and magnetic sectors of the theory.

Since $4D$ Euclidean invariance is broken by nonzero temperature the electric and
magnetic field strengths appear differently in the action. The magnetic field strength is
\be\label{B}
B^a_i=\frac{1}{2}\epsilon_{ijk}\left(\partial_jA_k^a-\partial_kA_j^a
+\epsilon^{abc}A_j^bA_k^c\right),
\ee
whereas the electric field strength consists of two pieces, the `static' and the `dynamical':
\be\label{E}
E_i^a=D_i^{ab}A_4^b-\dot A_i^a=\partial_iA_4^a+\epsilon^{acb}A_i^cA_4^b-\dot A_i^a.
\ee

In the $SU(2)$ gauge theory to which we mostly restrict ourselves in the present paper
there are only a few gauge and Euclidean $3D$ invariants in the order we
are interested in. These are $A_4^aA_4^a$, $E_i^aE_i^a$, $B_i^aB_i^a$ and
$(E_i^aA_4^a)^2$, $(B_i^aA_4^a)^2$. [For higher gauge groups there will be more
invariants.] The effective action (tree plus 1-loop) has the form
\bea
\label{effact}
S_{\rm eff}\!\! &=&\! \int\frac{d^3 x}{T} \left[- T^4 V(\nu)\! 
+\! E_i^2 F_1(\nu)\!+\! \frac{(E_i A_4)^2}{A_4^2} F_2(\nu)\!\right. 
+ \left. \! B_i^2\,H_1(\nu)\! +\!\frac{(B_i A_4)^2}{A_4^2}\,H_2(\nu)\! +\! \ldots\! \right]\,,\\
\nn \\
\nn
\nu &=&\frac{\sqrt{A_4^aA_4^a}}{2\pi T}.
\eea      
The static potential $V(\nu)$ has been known for 20 years \cite{GPY,Weiss};
the functions $F_{1,2},\,H_{1,2}$ are the new findings of this paper: they turn
out to be quite nontrivial and can be expressed through the digamma $\psi$ functions. 
The $E_i^2A_4^2$ and $B_i^2A_4^2$ terms of the effective action (corresponding
to the first terms of the Taylor expansion of our functions) have been known before
\cite{Chapman} and one combination (actually $F_1+F_2$ in our notations) was 
actually found previously by considering a particular case of $A_i=0$ \cite{BGK-AP}. We
agree with this previous work, however our results are, of course, more general. In addition to
the structures in \eq{effact} we have found a full-derivative term in the effective action. This
term is not necessarily zero: if the background field does not fall off fast enough at spatial
infinity it gives a finite contribution. This is, {\it e.g.}, the case when the background field is
that of the BPS dyon \cite{BPS}.    
  
Actually, quantum determinants are UV divergent, giving rise to the renormalization
of the bare coupling constant of the tree action. We perform an accurate regularization
of the determinants by means of the Pauli--Villars scheme. As a result, the above functions
are finite and the $F_1,\,H_1$ functions contain the running-coupling terms 
$\frac{11}{24\pi^2}\,\ln\left(\frac{T}{\Lambda}\,{\rm const.}\right)$ where $\Lambda$ is
the QCD scale in a particular regularization scheme. We have determined the value
of the `const.' in the argument of the logarithm and hence have learned the precise scale
of the running coupling constant at which it needs to be evaluated. Changing the
regularization scheme means the substitution $\Lambda_{\rm PV}=e^{\frac{1}{22}}
\Lambda_{\overline{\rm MS}} = 40.66\cdot\exp\left(-\frac{3\pi^2}{11 N^2}\right)\,
\Lambda_{\rm Lat}=\ldots$ \cite{Has}. 
 
There are two different approaches to the effective action and correspondingly two different
variants of the resulting functions $F_{1,2}...$ One can either exclude or include the
contribution of the static (zero Matsubara frequency) fluctuations to the effective action. One
follows the former logic if one wishes to get the effective action for static modes only. In this
case the potential energy $V(A_4)$ is not periodic and moreover it is formally UV divergent.
One follows the latter logic if one is interested, {\it e.g.}, in finding full quantum corrections to
semiclassical field configurations at nonzero temperatures, the examples of such being dyons \cite{dyon} and calorons \cite{caloron}. We compute the functions $F_{1,2}$ and $H_{1,2}$ in
both variants.   
 
Correspondingly, we think of two kinds of applications of our results. One is for studying the fluctuations and correlation functions of the Polyakov line in the region of temperatures where its average deviates considerably from the perturbative center-of-group values and where the dimensional reduction (i.e. perturbative) approximation fails. Another application is for evaluating the weights of semiclassical objects appearing at nonzero temperature \cite{D}.

\section{Basics of Yang-Mills theory at finite temperature}

The general definition of the partition function for statistical systems is
\bea 
{\cal Z} &=&  \sum_n\langle n\left| e^{-\beta {\cal H}}\right| n\rangle
\qquad\quad{\left[ \beta=\frac{1}{T} \right]}\\
 &=& \sum_n \int\!dq\, \psi^*_n(q)\; e^{-\beta {\cal E}_n}\; \psi_n(q)
=\int\!dq_0\int_{q(0)=q_0}^{q(\beta)=q_0}Dq(t)\,\exp\left(-\int_o^\beta\!dt{\cal H}[q,\dot q]
\right)\,,
\eea
where ${\cal H}$ is the Hamiltonian of the system, and ${\cal E}_n$ are its eigenvalues. In
Yang-Mills theory the role of coordinates $q$ is played by the amplitudes of the gluon fields
$A_i^a(x)$ and the Hamiltonian is
\be
\beta{\cal H}=\frac{1}{2g^2}\int_0^\beta\!dx_4\!\int\!d^3x\left[\left(\dot A_i^a\right)^2
+\left(B_i^a\right)^2\right]\,,
\label{Ham}\ee
where the dot indicates time derivative and $B_i^a$ is the magnetic field \ur{B}. 
The partition function can be written as a path integral over `trajectories'
$A_i(x_4, x)$ going from a `coordinate' $A_i^{(0)}$ at $x_4=0$ to the same
coordinate at $x_4=\beta$; one also has to integrate over this initial
coordinate:
\be\label{Zw}
{\cal Z} = \int DA_i^{(0)}(x) \int_{A_i(0,x)=A_i^{(0)}}^{A_i(\beta,x)=A_i^{(0)}} DA_i(x_4,
x)\,\exp\left\{-\frac{1}{2g^2}\,\int_0^\beta\!dx_4\, \int \,d^3x\left[\left(\dot A_i^a\right)^2
+\left(B_i^a\right)^2\right]\right\}\;.
\ee
However, in a gauge theory one sums not over all possible but only over
physical states, {\it i.e.} satisfying Gauss' law. In the absence of external
sources it means that only those states  need to be taken into account
that are invariant under gauge transformations:
\bea
A_i(x)\to \left[A_i(x)\right]^{\Omega(x)} &=&\Omega(x)^{\dagger}\,A_i(x)\,\Omega(x) +
i\Omega(x)^{\dagger}\,\partial_i \Omega(x) ,\\\nn
\qquad \Omega(x) &=& \exp\{i\,\omega_a(x)t^a\}\;.
\eea
To restrict the summation to physical states, one has to modify
eq. (\ref{Zw}). One projects to the physical {\it i.e.} gauge invariant states by
averaging the initial and final configurations over gauge rotations. 
The YM partition function is therefore
\bea\label{Zphys}
{\cal Z_{\rm phys}}\!\!\!&=&\!\!\sum_{\rm phys\; states}\langle n\left| e^{-\beta {\cal H}}\right|
n\rangle \nn \\
\!\!\!\!\!&=&\!\!\!\!\!\!\! \int \!\!\! D\Omega_{1,2}(x)D\!A_i^{(0)}(x) \!\!\!
\int_{A_i(0,x)=[A_i^{(0)}]^{\Omega_1(x)}}
^{A_i(\beta,x)=[A_i^{(0)}]^{\Omega_2(x)}} \!\!\!\!
D\!A_i(x_4,x)\,\exp\!\left\{\!\!-\frac{1}{2g^2}\!\! \int_0^\beta\!\!\!dx_4\!\! \int \!\!\!
  d^3x \!\!\left[\!\left(\!\dot A_i^a\!\right)^2
\!\!\! + \!\!\left(B_i^a\right)^2\right]\!\right\}\!.
\eea
Renaming the initial field $[A_i^{(0)}]^{\Omega_1(x)} \to
A_i^{(0)}$ and introducing the relative gauge transformation $\Omega(x)=\Omega_2(x)\,
\Omega_1^{\dagger}(x)$ one can rewrite this as  \cite{GPY}
\be\label{Zph1}
{\cal Z_{\rm{phys}}}=\, \int \, D\Omega(x)DA_i^{(0)}(x)
\int_{A_i^{(0)}}^{[A_i^{(0)}]^{\Omega(x)}}\, 
DA_i(x_4,x)\;e^{-\beta{\cal H}[A_i(x_4,x)]}\;.
\ee
There is a subtle question whether one has to include integration over global
gauge transformations, {\it i.e.} $x$-independent $\Omega$'s in eq. (\ref{Zph1}). If
one does, it means that only states with total color charge zero are admitted
in the partition function. A more cautious approach is to allow for states
with nonzero color charge: if these are for some reasons dynamically
suppressed it must be seen from the theory but not imposed by
hand. Therefore we shall admit x-independent $\Omega$'s but not integrate
over them explicitly. 

In order to put the partition function into a more customary four dimensional
form one introduces an interpolating gauge transformation $\Omega(x_4,x)$
such that
\bea
\Omega(x_4,x)&=&\left\{\begin{array}{cc}{\bf 1},& x_4=0,\\ \Omega(x),& x_4 =\beta
\end{array}\right. \;. 
\eea
Simultaneously one changes the integration variables from $A_i(x_4,x)$ to 
\be
A_i^\prime(x_4
,x)=\Omega(x_4,x)\,A_i(x_4,x)\Omega^\dag(x_4,x)+i\Omega(x_4,x)\,\partial_i\,
\Omega^\dag(x_4,x)
\ee
and introduces, instead of $\Omega(x_4,x)$,  the new variable
\be
A_4(x_4,x)=i\Omega(x_4,x)\,\partial_{4}\,\Omega^\dag(x_4,x).
\ee
For example, if the interpolating gauge transformation is taken to be $\Omega(x_4
,x)=\exp\{i\,x_4\,T\,\omega^a(x)\,t^a\}$, then $A_4$ is time-independent and
equal to $A_4(x)= T\,\omega^a(x)\,t^a$. We note that both $A_4(x_4,x)$ and
$A_i^\prime(x_4 ,x)$ are periodic in temporal direction. 

The magnetic energy is gauge invariant, {\it i.e.} 
\be
\Tr\,B^2(A_i) = \Tr\,B^2(A_i^\prime)\;,
\ee
while the electric energy becomes 
\be
\Tr\,E^2 = \Tr \,\dot A_i^2 = \Tr\,{E^\prime}^2\;,
\ee
where
\be
E_i^\prime = \dot A_i^\prime - \partial_i\,A_4 - i\,[A_4, A_i^\prime]
=\dot A_i^\prime-[\nabla_i(A^\prime)A_4]. 
\ee
Therefore the full action density can be rewritten as a standard
$\Tr\,F_{\mu\nu}^2$, where
\be
F_{\mu\nu}=\partial_{\mu} A_{\nu} - \partial_{\nu}A_{\mu}
-i\,[A_{\mu},A_{\nu}]
\ee
with $A_{\mu}(x_4,x)$ denoting $A_i^\prime(x_4,x)$ and $A_4(x_4,x)$. Thus,
eq. (\ref{Zph1}) is equivalent to the more familiar partition function
\be\label{Zph2}
{\cal Z}_{\rm phys}=\int
DA_\mu\,\exp\left\{-\frac{1}{4g^2}\int\!d^4x\,F_{\mu\nu}^aF_{\mu\nu}^a\right\},
\ee
where one integrates over gauge fields obeying periodic boundary conditions in
time, meaning $A_{\mu}(x_4, x) = A_{\mu}(x_4 + \beta, x)$,
with $\beta = 1/T$.

Periodic fields can be decomposed into Fourier modes: 
\be
A_\mu(x_4,x)=\sum_{k=-\infty}^\infty A(\ok,x)\,e^{i\ok x_4},
\qquad {\ok=2\pi k\,T.}\;
\ee
where $\ok=2\pi k\,T$ are the so-called Matsubara frequencies, which play the
role of mass. In the limit $T\to\infty$ all nonzero Matsubara modes become infinitely
heavy. If one leaves only the static gluon modes it is called {\it dimensional reduction}
\cite{DR}, as the resulting theory is purely static. There is no dynamics in the time direction
anymore. At high, but not infinite temperatures, this approximation is too crude. The nonzero
modes show up in loops and produce infinitely many effective vertices. The aim of this paper
is to find all these infinite number of vertices restricted, however, to low momenta $p<T$,
induced in the 1-loop order. 

\section{One loop quantum action}

As stressed in the Introduction, one can always choose the background field $A_4$
to be static. As to the $A_i$ field, we shall temporarily {\em take} it to be static: the 
generalization of the effective action to the case of timedependent $A_i$ will be simple.

To study the effects of the nonzero Matsubara modes we use a background field method
and split the gluon fields into a time independent background field $\bar{A}_{\mu}(x)$ 
and a presumably small quantum fluctuation field $a_{\mu}(x_4,x)$:
\be\label{bfgl}
A_{\mu}(x_4, x) = \bar{A}_{\mu}(x) + a_{\mu}(x_4, x)\;.
\ee
In this paper we consider the quantum effects at the 1-loop level. Then it is
sufficient to expand the action around the background field up to quadratic order in
$a_{\mu}$. The linear term in $a_\mu$ is absent owing to the orthogonality of nonstatic
modes to static ones. We shall, however, also investigate the contribution of the
static fluctuation mode. In this case the linear term is absent if, {\it e.g.}, the background
field satisfies the equation of motion or if the static mode is varying in space faster
than the background field. The quadratic form is, generally speaking, degenerate so that one
has to fix the gauge for fluctuations. This gauge fixing is unrelated to the gauge fixing
of the background field. We choose the background Lorenz gauge
$D_{\mu}(\bar{A})a_{\mu}=0$ \footnote{Jackson and Okun \cite{JO} recommend
to name the $\partial_\mu A_\mu=0$ gauge after the Dane Ludvig Lorenz
and not after the Dutchman Hendrik Lorentz who certainly used this gauge too 
but several decades later.}, where
\be
D_\mu^{ab}(\bar{A}) = \partial_{\mu}\delta^{ab} + f^{acb}\bar{A}_{\mu}^{c}\;.
\ee
is the covariant derivative in the adjoint representation. 
This gauge brings in the Faddeev-Popov ghost determinant which can be expressed as a
Grassmann integral over ghost fields. For the partition function this yields
\be\label{pfbf}
Z(\bar{A}) = e^{S} = e^{\bar{S}}\, \int{}Da\,D\chi\,D\chi^{+}\,\exp \left\{-
\frac{1}{2g^2(M)}\int{}d^4 x\,a_{\mu}^b\, W_{\mu\nu}^{bc}\, a_{\nu}^c - \int \, d^4
x\,\chi^{+a}\left(D^2_{\mu}\right)\,\chi^{a}\right\}\;,
\ee
where $\chi, \chi^+$ are ghost fields and 
\be
 \bar{S} =  -\frac{1}{4 g^2(M)}\int\,d^4 x{}F_{\mu\nu}^{a}(\bar{A}) F_{\mu\nu}^{a}(\bar{A}) 
\ee
is the action of the background field. The quadratic form for $a_{\mu}$ in the background
Lorenz gauge is given by
\be\label{W}
W_{\mu\nu}^{ab} = -[ D^2(\bar{A})]^{ab} \delta_{\mu\nu} - 2 f^{acb}F_{\mu\nu}^c(\bar{A}) \;.
\ee
Integrating out the quantum fluctuations and ghosts yields two functional determinants,
\be\label{zf}
Z(\bar{A}) = e^{\bar{S}}\; \left(\det{W}\right)^{-1/2} \; \det\left( -D^2\right)\;,
\ee
so that the 1-loop action is
\be\label{Seff}
S_{\rm{ 1-loop}} = \log\, \left(\det{W}\right)^{-1/2} + \log\,\det \left(-D^2\right)\;.
\ee

Since the operators $D^2, W_{\mu\nu}$ are built from covariant derivatives and the field
strength only, this action is invariant under general gauge transformations of the background
field. One can use this freedom to make the $A_4$ component static, which we shall always
assume. The spatial components $A_i$ are then, generally speaking, time dependent. For the
most of the paper we shall assume that $A_i$ is time-independent too. At the end we shall be
able to reconstruct terms with $\dot A_i$ from gauge invariance but at the time being we
shall take static $A_i$. Then the quantum action \ur{Seff} is invariant under time-independent
gauge transformations,
\bea
\bar{A_4}(x) &\to & U(x)\,\bar{A_4}(x)\,U^\dag(x),\\
\bar{A_i}(x) &\to & U(x)\,\bar{A_i}(x)\,U^\dag(x)+i\,U(x)\,\partial_i\,U^\dag(x)\;.
\eea
In this paper we restrict ourselves to the $SU(2)$ color group, which means that the action
depends on the gauge and $3D$ Euclidean invariants $A_4^a A_4^a$, $E_i^a
E_i^a$, $B_i^a B_i^a$, $E_i^a A_4^a$, $B_i^a A_4^a$, etc. For higher groups
there will be more invariants. We write the background fields without a bar from now on, as
they are the only field variables left. 

In fact the action can be presented as a series in the spatial covariant derivative $D_i$.
Since the electric field is given by $E_i^{a} = D_i^{ab} A_4^b$, an expansion in powers of
the electric field corresponds to a covariant gradient expansion of the $A_4$ fields. To get
the magnetic field, we already need one more power of $D_i$, as
$B_k^a=\frac{1}{2}\epsilon_{ijk}F_{ij}^a=\frac{1}{4}\epsilon_{ijk}\epsilon^{cad}\left[D_i, D_j\right]^{cd}$. 
For the $SU(2)$ gauge group, in the electric (magnetic) sector only two independent color
vectors exist, $E_i^a$ ($B_i^a$) and $A_4^a$. Therefore, we expect the following structure
for the gauge-invariant gradient expansion:
\bea\label{proph}
S_{\rm{ 1-loop}}\!\! =\! \int\frac{d^3 x}{T} \left[-T^4\,V(\nu)\! +\!
  E_i^2 f_1(\nu)\! +\! \frac{(E_i A_4)^2}{A_4^2} f_2(\nu)\! +\! B_i^2\,h_1(\nu)\! +\!
\frac{(B_i A_4)^2}{A_4^2}\,h_2(\nu)\! +\! \ldots\! \right].
\eea
In the explicit evaluation of the functional determinants we find exactly the
structure eq. (\ref{proph}) and determine the functions $f_1, f_2, h_1, h_2$ at
all values of their argument which is in fact a dimensionless ratio
$\nu=\sqrt{A_4^a A_4^a}/(2\pi T)$. 

\section{The functional determinants}

We start with the evaluation of the ghost functional determinant. As usual we
subtract the zero gluon field contribution. Using the fact that 
$\det\,{\rm  K}=\exp\,\Tr\,\log\,{\rm K}$ we can write
\be\label{ghost}
\exp\,\log\,\frac{\det (-D_{\mu}^2)}{\det (-\partial_{\mu}^2)} = \exp\;\Sp\,[\log\,(-D_{\mu}^2) -
\log\,(-\partial_{\mu}^2)] 
\ee
where $\Sp$ is a functional trace. We present the ratio of determinants with the help
of the Schwinger proper time representation \cite{Schw}: 
\be\label{logdet}
\log\,\det (-D^2)_{\rm{n}} \equiv \log\,\frac{\det (-D_{\mu}^2)}{\det (-\partial_{\mu}^2)} =
-\int_0^{\infty}\frac{ds}{s}\;  \Sp \left( e^{s D_{\mu}^2} - e^{s \partial_{\mu}^2} \right)\;.
\ee
In fact this ratio is logarithmically UV divergent, reflecting the coupling constant
renormalization. We use the Pauli-Villars method to regularize the divergence. This
corresponds to replacing the determinant by a `quadrupole formula':
\bea
\lefteqn{\det (-D^2)\longrightarrow \det (-D^2)_{\rm{r, n}}} \\
\equiv\frac{\det (-D_{\mu}^2)}{\det (-\partial_{\mu}^2)}\,\frac{\det
 (-\partial_{\mu}^2 + M^2)}{\det (-D_{\mu}^2 + M^2)} &=& \exp\left\{
-\int_0^{\infty}\frac{ds}{s}\;  \Sp \left[\left( 1 - e^{-s M^2}\right) \,\left( e^{s D_{\mu}^2} 
- e^{s \partial_{\mu}^2} \right)\right]\right\}\label{pv}\;.
\eea
The functional trace in eq. (\ref{pv}) can be taken by inserting any full
basis, so we are free to choose {\it e.g.} the plane-wave basis, $\exp (ix_{\alpha} p_{\alpha})$. 
Then, by the definition of the functional trace, one can write
\be
{\rm Sp}\, e^{-s K} = \Tr\,\int\,d^4 x\,\lim_{y\to x}\int\frac{d^4 p}{(2 \pi)^4}\,
\exp(-ip\cdot y)\exp(-s K)\,\exp(ip\cdot x),
\label{Spdef}\ee 
where $\Tr$ is the remaining matrix trace over color and, as the case may be, Lorentz
indices. One can now drag the latter plane-wave exponent though the differential operator
$K$ until it cancels with the former. This results in the shift of the derivatives inside the
differential operator and in the following representation of the functional trace \cite{DPY}:
\be\label{shift}
{\rm Sp}\, e^{-s K} = \Tr\,\int\,d^4 x\,\int\frac{d^4 p}{(2 \pi)^4}\,\exp \left[-s K (\partial_{\alpha}
\rightarrow \partial_{\alpha}+ i p_{\alpha})\right]\bf{1}\,.
\ee
The $\bf{1}$ at the end is meant to emphasize that the shifted operator acts on unity,
so that for example any term that has a $\partial_{\alpha}$ in the exponent and is brought all the
way to the right, will vanish. According to ({\ref{shift}) we now have
\bea\label{gh2}
\lefteqn{\log\,\det (-D^2)_{\rm{r,n}}
=
- \int\,\frac{d^3 x}{T}\,T \sum_{k=-\infty}^{\infty}\int\frac{d^3 p}{(2
\pi)^3}\,\int_0^{\infty}\frac{ds}{s} \left( 1 - e^{-s M^2}\right)}\\ & \times& \Tr\,\left\{\exp\left[s
(D_4 + i \ok)^2 + s ( D_i + i p_i)^2\right] - \exp\left[s (i \ok)^2 + s (i p_i)^2\right]\right\}\;.
\eea
Owing to the periodic boundary conditions we have replaced the integration over
$p_4$ by the sum over the Matsubara frequencies $\omega_k = 2 \pi k T$ and
taken into account that the $x_4$ integration goes from $0$ to $\beta=1/T$. Keeping in mind
that the background field is time independent one can replace
\be
D_4^{ab} \to f^{acb} A_4^c\;.
\ee
We define the adjoint matrix
\be
\A^{ab} = f^{acb} A_4^c + i \omega_k \delta^{ab}
\ee
upon which eq. (\ref{gh2}) becomes
\bea\label{gh3}
\lefteqn{\log\,\det (-D^2)_{\rm{r,n}}
= - \int\,d^3 x\,\sum_{k=-\infty}^{\infty}\int\frac{d^3 p}{(2
  \pi)^3}\,\int_0^{\infty}\frac{ds}{s}\,\left( 1 - e^{-s M^2}\right)
}\\ &\times&
\Tr\,\left\{\exp\left[ s \A^2 + s D_i^2 + 2isp_iD_i - sp^2\right] -  \exp\left[-s(\ok^2 +
p^2)\right]\right\} \;.
\eea

In the same way as for the ghost determinant (\ref{ghost}) we use the
`quadrupole formula' and write the normalized and regularized gluon determinant as 
\be
\log\,(\det W)^{-1/2}_{{\rm r,n}} =
\frac{1}{2}\,\int_0^{\infty}\frac{ds}{s}\left( 1 - e^{-s M^2}\right)\Sp \left(e^{-s W_{\mu\nu}^{ab}} -
e^{s \partial^2\,\delta_{\mu\nu}\delta^{ab}}\right)\;,
\ee
which after an insertion of a plane wave basis and dragging $\exp(ip\cdot x)$ through
the differential operator yields
\bea\label{gluon}
\lefteqn{\log\,(\det W)^{-1/2}_{{\rm r,n}} = 
\frac{1}{2}\int\,d^3 x\,\sum_{k=-\infty}^{\infty}\int\frac{d^3 p}{(2
 \pi)^3}\,\int_0^{\infty}\frac{ds}{s}\left( 1 - e^{-s M^2}\right)}
\\&\times&
\Tr\,\left\{\exp\left[(s \A^2 + s D_i^2 + 2isp_iD_i - sp^2)^{ab} \delta_{\mu\nu} +
2sf^{acb}F_{\mu\nu}^c\right]
-  \exp\left[-s(\ok^2+p^2)\right]\right\}\;.\nn
\eea
A covariant gradient or derivative expansion is the expansion in $D_i$, applied to
$\A^2$ and $F_{\mu\nu}^c$. For example, to quadratic order in $D_i$ it corresponds to
summing up all 1-loop Feynman diagrams with two $A_4$ vertices carrying momenta,
and any number of $A_4$ insertions at zero momentum. So far, both eq. (\ref{gh3}) and
eq. (\ref{gluon}) are independent of the gauge group. 

\section{Zeroth order of covariant derivative expansion}

Zeroth order in the expansion corresponds to setting $D_i=0$. For the gluon
part (\ref{gluon}) the field strength does not contribute at this order since
it is quadratic in the covariant derivatives. Hence the gluonic contribution
is $-(1/2)\times 4 = -2$ times the ghost contribution, where the factor
$(-1/2)$ comes from the fact that the gluon determinant is taken to this
power  (see \ur{zf}), and the Lorentz structure of the gluons, $\Tr\,\delta_{\mu\nu}$, yields 4. 
At zeroth order of the covariant derivative expansion one thus has:
\be
S_{\rm{ 1-loop}}^{(0)} = -\left[\log\,\det \left(-D^2\right)\right]^{(0)}\;.
\ee
The determinant is UV finite in this order, so one does not need to regularize it. For the
explicit calculation one can choose a gauge where $A_4$ is diagonal in the fundamental
representation, hence for the $SU(2)$ gauge group
\be\label{gauge}
A_4^a = \delta^{a3}\phi \qquad,\qquad \phi = \sqrt{A_4^a\,A_4^a}\;.
\ee
The eigenvalues of the $3\times 3$ matrix $\A^{ab} = \epsilon^{a3b}\phi + i \ok\delta^{ab}$ are
$\left(\ok+\phi, \ok-\phi, \ok \right)$. It is obvious that upon summation over all
Matsubara frequencies $f[(\ok + \phi)^2]$ and $f[(\ok - \phi)^2]$ give the
same. We hence obtain 
\be
S_{\rm{ 1-loop}}^{(0)} =  2 \int\,d^3 x\,\sum_{k=-\infty}^{\infty}\int\frac{d^3 p}{(2
\pi)^3}\,\int_0^{\infty}\,\frac{ds}{s}\,e^{-s p^2}\left\{e^{-s (\ok - \phi)^2} - e^{-s \ok^2}\right\}\;.
\ee
Integrating over proper time $s$ and summing over Matsubara frequencies labeled by $k$
gives
\be\label{key1}
\sum_{k=-\infty}^{\infty}\int_0^{\infty}\,\frac{ds}{s}\,e^{-s (\ok - \phi)^2 - s p^2} =
-\log\left(\ch\frac{|\vec{p}|}{T} - \cos\frac{\phi}{T}\right)\;.
\ee
The $p$ integration can be performed with the help of Ref. \cite{GR}. One obtains
\be\label{S01loop}
S_{\rm{ 1-loop}}^{(0)} =  -\int d^3 x \frac{1}{12 \pi^2 T}\phi^2\left( 2\pi T - \phi\right)^2{ |
}_{{\rm mod}\; 2\pi T} \;,
\ee
hence the dimensionless static potential is
\be\label{0thf}
V(\nu) =  \frac{(2\pi)^2}{3}\,\nu^2\,(1-\nu)^2{ |}_{{\rm mod} \;1},  
\qquad \nu=\frac{\sqrt{A_4^a A_4^a}}{2\pi T}\,.
\ee
This result is well known \cite{GPY, Weiss}. We want to stress here that the term cubic in
$\nu$ arises solely from the zero Matsubara frequency, $k=0$. It makes eq. (\ref{S01loop})
periodic in $\nu$ with unit period. It should be noted that without the zero frequency
contribution the $p$ integration is UV divergent; the addition of the $\ok=0$ mode removes
this divergence. 
  
\section{General technique for the covariant derivative expansion}

In the next orders in the covariant derivative the calculation becomes more
involved. We wish to keep all powers of $A_4$, but expand in powers of $D_i$.
To expand the exponential of two noncommuting operators A and B we use the formulae
 \be\label{master1}
e^{A+B} = e^A + \int_0^1\,d\alpha\,e^{\alpha A}B e^{(1-\alpha) A} + \int_0^1\,d\alpha\,\int_0^{1-\alpha}\,d\beta\,e^{\alpha
A}B\,e^{\beta A} B e^{(1-\alpha-\beta)}A + \ldots \;
\ee
and
\be\label{master2}
\left[ B, e^A\right] = \int_0^1 d\gamma\,e^{\gamma A}\left[ B,A \right]\,e^{(1-\gamma)A}\;.
\ee
Here $B$ denotes the combination of covariant derivatives in the exponents in \eq{gh3} or
\eq{gluon}, $A$ is everything that is left there. We encounter here the following commutators:
 \bea\label{cDa}
\left[D_i, \A\right] &=& \left[D_i,D_4\right] = -i\,F_{i4} = -i E_i\;, \\
\label{cDaa}
\left[D_i, \A^2\right] &=& -i \left\{\A, E_i\right\}\;,
\eea
where the electric field is in the adjoint representation, {\it i.e.}
\be\label{ei}
E_i^{ab} = i f^{acb}E_i^c\;.
\ee
The strategy is to drag all derivatives in $B$ to the right using the master formulae 
\urs{master1}{master2}.

\section{Electric sector}

We are now going to find the second and third terms in eq. (\ref{proph}),
{\it i.e.} terms quadratic in the covariant derivative $D_i$. As expected, terms
which are not gauge invariant cancel out individually for the ghosts and the
gluons. In the end three different gauge invariant contributions
remain. Writing down the ghost determinant as
\be\label{Dexp}
\log\,\det (-D^2)_{\rm{r,n}}
= - \int\,d^3 x\,\sum_{k=-\infty}^{\infty}\int\frac{d^3 p}{(2\pi)^3}\,
\int_0^{\infty}\frac{ds}{s}\left( 1 - e^{-s M^2}\right) \Tr\,e^{-sp^2}\left[e^{A+ B}\right] \;,  
\ee 
where $A = s \A^2$ and $B = 2isp_iD_i + s D_i^2$ and expanding in $B$ with
the use of eqs. (\ref{master1},\ref{master2}) we obtain [see Appendix B.1]
\be\label{logdetD}
\left[\log\,\det (-D^2)_{{\rm r,n}}\right]_{{\rm E}}^{(2)}
= - \int\,d^3 x\,\sum_{k=-\infty}^{\infty}\int\frac{d^3 p}{(2\pi)^3}\,
\int_0^{\infty}\frac{ds}{s}\left( 1 - e^{-s M^2}\right) \,e^{-s p^2}\left[I_1 + I_2\right]\;.
\ee
The gluon determinant is
\be\label{Wexp}
\log\,(\det W)^{-1/2}_{{\rm r,n}} = \frac{1}{2}\int\,d^3
x\,\sum_{k=-\infty}^{\infty}\int\frac{d^3 p}{(2
  \pi)^3}\,\int_0^{\infty}\frac{ds}{s}\left( 1 - e^{-s M^2}\right)
\Tr\,e^{-sp^2\delta_{\mu\nu}}\left[e^{A+B}\right]\;,
\ee
where this time $A_{\mu\nu}^{ab} = (s \A^2)^{ab}\,\delta_{\mu\nu}$ and
$B_{\mu\nu}^{ab} = (s D_i^2 + 2isp_iD_i)^{ab} \delta_{\mu\nu} +
2sf^{acb}F_{\mu\nu}^c$. Expanding in $B_{\mu\nu}^{ab}$ and using
eqs. (\ref{master1},\ref{master2}) we find [see Appendix B.1]
\be\label{logdetW}
\left[\log\,(\det W)^{-1/2}_{{\rm r,n}}\right]_{{\rm E}}^{(2)} = \int\,d^3
x\,\sum_{k=-\infty}^{\infty}\int\frac{d^3 p}{(2
  \pi)^3}\,\int_0^{\infty}\frac{ds}{s} \left( 1 - e^{-s M^2}\right)\,e^{-s p^2}\left[2 I_1 + 2 I_2 +
\frac{I_3}{2}\right]\;.
\ee
The gauge invariants in eqs. (\ref{logdetD},\ref{logdetW}) are
\bea\label{I10}
I_1 &=& \!\!s^3\!\int_0^1 \!d\alpha \! \left\{\!-\frac{1}{2} + \alpha(1\!-\!\alpha)+\frac{2}{9}s
p^2\left[1\!-\!\frac{3}{2}\alpha(1\!-\!\alpha)\right]\!\right\}\!\! \Tr\,e^{(1-\alpha) s\A^2}\!\left\{\A, E_i\right\} \!
e^{\alpha s\A^2} \!\!\left\{\A, E_i\right\}\!,\\ 
\label{I20} 
I_2 &=& \!\!-s^2\left( \frac{1}{2}-\frac{2}{9}sp^2 \right)\, \Tr\,e^{s\A^2}\left(2 E_i^2 +
i\left\{\A,\left[D_i, E_i\right]\right\}\right),\\
\label{I30}
I_3 &=& \!\!8s^2\int_0^1\,d\alpha\,\frac{1}{2}\,\Tr\,e^{(1-\alpha) s\A^2}\,E_i\,e^{\alpha s\A^2}\,E_i\,. 
\eea
The total 1-loop action (\ref{Seff}) is
\bea\label{Sn}
\lefteqn{\left[S_{\rm{1-loop}}^{(2)}\right]_{{\rm E}} = \left[\log\, \left(\det{W}\right)^{-1/2}_{{\rm
n}} +
  \log\,\det \left({-D^2}\right)_{{\rm r,n}}\right]^{(2)}_{{\rm E}}}\\ \nn &=&
\int\,d^3x\,\sum_{k=-\infty}^{\infty}\int\frac{d^3 p}{(2
  \pi)^3}\,\int_0^{\infty}\frac{ds}{s} \left( 1 - e^{-s M^2}\right)\,e^{-s p^2}\left[I_1 + I_2 +
\frac{I_3}{2}\right]\;.
\eea
For the explicit evaluation we have to do all the integrations over $\alpha$, $s$, $p$ and the
summation over the Matsubara frequencies $\ok$. For convenience we rescale the field
variable and introduce
\be
\phi = 2 \pi T \nu 
\qquad {{\rm where}} \qquad
0\le \nu \le 1 \;.
\ee
The case when $\nu$ is outside this interval will be considered separately.
In the invariant $I_2$ we will here only take the first term and leave away the anticommutator.
Its effect will be shown later. 
In the sum over the Matsubara frequencies we treat the zero mode separately
[see Appendix B.2 for details]. For $I_1$ and the first term in $I_2$ the zero Matsubara
frequency
yields each an IR divergent term {\it i.e.} proportional to $\lim_{\ok \to 0}(1/\ok)$
and one finite term which is proportional to $1/\phi$. The `naked' IR divergencies cancel
between the two invariants $I_{1,2}$ in such a way that both the ghost and the gluon
contribution are separately IR finite. The zero mode of the invariant $I_3$ contributes only
with a finite $1/\phi$ term. 

With our gauge choice $A_4^a=\delta^{a3}\phi$ we actually get two structures, 
$(E_i^1E_i^1 + E_i^2E_i^2) f_1(\phi)$ and $E_i^3 E_i^3 f_3(\phi)$, which can be written in
invariant terms as 
\be
E_i^2 f_1(\nu) \qquad {\rm and}\qquad \frac{(E_i\cdot A_4)^2}{|A_4|^2}
\left( f_3(\nu)-f_1(\nu)\right)\;,
\ee
respectively. Adding up the ghost and the gluon result and denoting $(f_3-f_1)$ by $f_2$ we
find the two functions defined in eq.(\ref{proph}):
\bea
f_1 &=& \frac{11}{48 \pi^2}\left[2 \left( \log\,\mu - \ge\right) - \psi\left(-\frac{\nu}{2}\right) -
\psi\left(\frac{\nu}{2}\right) + \frac{20}{11\nu}\right] \label{f1}\;, \\
f_2 &=& \frac{11}{48 \pi^2}\left[\psi\left(-\frac{\nu}{2}\right) +
  \psi\left(\frac{\nu}{2}\right) -\psi\left(\nu\right) -
  \psi\left(1-\nu\right) - \frac{20}{11\nu}\right] \label{f2}\,,\\
f_3 &=&  \frac{11}{48 \pi^2}\left[2 \left( \log\,\mu - \ge\right) -
  \psi\left(\nu\right) - \psi\left(1-\nu\right)\right]\,,
\qquad\nu=\frac{\sqrt{A_4^aA_4^a}}{2\pi T} \;.
 \label{f3}
\eea
Here $\psi$ is the digamma function,
\be
\psi(z) = \frac{\partial}{\partial\,z}\log\,\Gamma(z)\;,
\ee
$\ge$ is the Euler constant, and the argument of the logarithm $\mu$ is the cutoff
that we have introduced in the sum over Matsubara frequencies [see Appendix B.2]. 
It is related to the Pauli-Villars mass as
\be\label{mu}
\mu= \frac{M}{4\pi T}\,e^{\ge}\,.
\ee
This result for the Pauli-Villars scheme agrees with Ref. \cite{coupling} where the scale of the running coupling constant in the dimensionally reduced theory was  studied in the ${\overline{\rm MS}}$ scheme.  The subtraction scales are related according to Ref. \cite{Has}.

Using eq.(\ref{mu}) we can express the functions $f_{1,2,3}$ as:
\bea
\label{f11}
f_1 &=& \frac{11}{48 \pi^2}\left[2 \log\,\frac{M}{4\pi T} -
\psi\left(-\frac{\nu}{2}\right) - \psi\left(\frac{\nu}{2}\right) + \frac{20}{11\nu}\right] \;, \\
\label{f21}
f_2 &=& \frac{11}{48 \pi^2}\left[\psi\left(-\frac{\nu}{2}\right) +
  \psi\left(\frac{\nu}{2}\right) -\psi\left(\nu\right) -
  \psi\left(1-\nu\right) - \frac{20}{11\nu}\right]\;,\\
\label{f31}
f_3 &=&  \frac{11}{48 \pi^2}\left[2 \log\,\frac{M}{4\pi T}  -
  \psi\left(\nu\right) - \psi\left(1-\nu\right)\right]\;. 
\eea

Recalling that the tree-level action has the bare coupling defined at the
cutoff momentum $M$,
\be
\frac{E_i^a E_i^a}{2 g^2(M)} = E_i^a\,E_i^a\, \frac{11}{24 \pi^2}\,\log\frac{M}{\Lambda}\,,
\ee
we see that the UV divergent $\log\,M$ term cancels out in the sum of tree and 1-loop
actions. The full action is finite and can be presented in the form of \eq{effact} where
the functions $F_{1,2,3}$ are obtained from $f_{1,2,3}$ by replacing the cutoff $M$ by the
finite $\Lambda$ parameter:
\bea
\label{F1}
F_1 &=& \frac{11}{48 \pi^2}\left[2 \log\,\frac{\Lambda}{4\pi T}  -
\psi\left(-\frac{\nu}{2}\right) - \psi\left(\frac{\nu}{2}\right) + \frac{20}{11\nu}\right] \;, \\
\label{F2}
F_2 &=& \frac{11}{48 \pi^2}\left[\psi\left(-\frac{\nu}{2}\right) +
  \psi\left(\frac{\nu}{2}\right) -\psi\left(\nu\right) -
  \psi\left(1-\nu\right) - \frac{20}{11\nu}\right]\;, \\
\label{F3}
F_3 &=&  \frac{11}{48 \pi^2}\left[2 \log\,\frac{\Lambda}{4\pi T}  -
  \psi\left(\nu\right) - \psi\left(1-\nu\right)\right],\qquad\nu=\frac{\sqrt{A_4^aA_4^a}}{2\pi T}.
\eea
The functions $F_1$ and $F_3$ without the first term are plotted in Fig.\ref{L1-L3}. 
Both functions are singular at $A_4\to 0$, which is due to the contribution of the zero
Matsubara frequency. 

\begin{figure}[t]
\centerline{
\epsfxsize=0.45\textwidth
\epsfbox{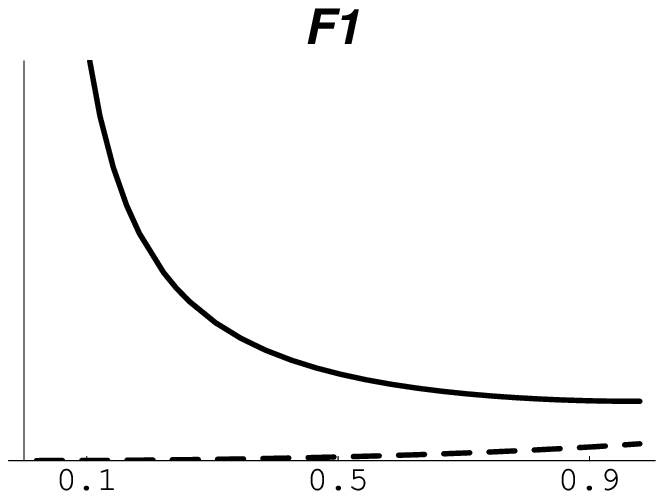}
\hspace{5mm}
\epsfxsize=0.45\textwidth
\epsfbox{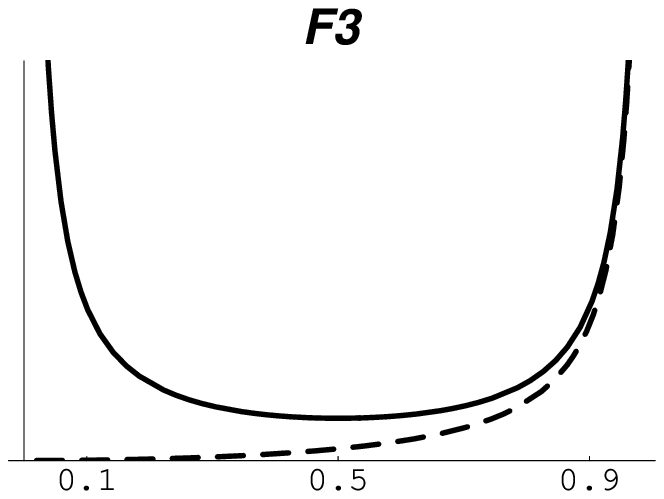}}
\caption{ {\small $F_1(\nu)$ (left) and $F_3(\nu)$ (right) without the constant first terms are
shown for $0\le \nu \le 1$. The solid lines include the contribution of the zero Matsubara
frequency, while it is subtracted in the dashed curves}.\label{L1-L3} }
\end{figure}

\subsubsection{Electric sector without the zero Matsubara frequency contribution}

The functions $F_1$ and $F_2$ have so far been evaluated by summing over all
the Matsubara frequencies, including the static fluctuations
around static gluon background fields. This is of interest for a number of
physical cases. For the problem of dimensional reduction, however, one does not
include the static quantum fluctuations. Subtracting the contributions
of the zero Matsubara frequency, which are of order $1/\nu$, we obtain
\bea
\label{F1t}
\tilde{F}_1(\nu) &=& \frac{11}{48 \pi^2}\left[2 \log\,\frac{\Lambda}{4\pi T}  -
\psi\left(-\frac{\nu}{2}\right) - \psi\left(\frac{\nu}{2}\right) \right]\;, \\ 
\label{F2t}
\tilde{F}_2(\nu) &=& \frac{11}{48 \pi^2}\left[\psi\left(-\frac{\nu}{2}\right) 
+ \psi\left(\frac{\nu}{2}\right) -
\psi\left(\nu\right) - \psi\left(-\nu\right)\right] \;, \\ 
\label{F3t}
\tilde{F}_3(\nu) &=& \frac{11}{48 \pi^2}\left[2 \log\,\frac{\Lambda}{4 \pi T}  -
\psi\left(\nu\right) - \psi\left(-\nu\right)\right]\,. 
\eea
Here we used the relation for the $\psi$ function: $\psi(1+\nu)=\psi(\nu)+\frac{1}{\nu}$.
Keeping in mind that
\be\label{psiexpan}
-\po(x)-\po(-x)=2\left[\gamma_E+\zeta(3)x^2+\zeta(5)x^4+\ldots\right]
\ee
where $\zeta(n)$ is the Riemann zeta function, we see that the contribution of the nonzero
Matsubara frequencies to the effective action is regular at $A_4=0$, and that the bare
coupling constant should be replaced by the running one taken at the scale 
$4\pi e^{-\gamma_E}\,T\approx \!7.05551\,T$, if the Pauli-Villars $\Lambda$ is used. If
another regularization scheme is used, the scale should be changed accordingly, see the
Introduction.

The plots of $\tilde F_{1,3}$ (without the first terms) are also shown in Fig.\ref{L1-L3}. 

\subsubsection{The `equation of motion' term}

We finally compute the second term in the invariant $I_2$, which we have so far left out. 
Its contribution  is
\be\label{eof}
S_{{\rm EoM}} = \frac{1}{T}\int\,d^3 x\,\sum_{k=-\infty}^{\infty}\int\frac{d^3 p}{(2
\pi)^3}\,\int_0^{\infty}\frac{ds}{s}\,e^{-s p^2}\,T\,\left[-s^2\left( \frac{1}{2}-\frac{2}{9}sp^2 \right)\,
\Tr\,e^{s\A^2}\left(i\left\{\A,\left[D_i, E_i\right]\right\}\right)\right]\;,
\ee
where the curly brackets denote the anticommutator. After all integrations and the summation
over $\ok$ it gives
\be\label{eom}
S_{{\rm EoM}} = \int\,d^3 x\,\frac{1}{12\pi}\left[\frac{(D_i E_i)^a\,A_4^a}{\pi T} - \frac{(D_i
E_i)^b\,A_4^b}{\sqrt{A_4^c\,A_4^c}}\right]\;.
\ee
Here the first term comes exclusively from the nonzero Matsubara
frequencies, while the second term is the contribution of the zero Matsubara
frequency alone.  Equation (\ref{eom}) is zero if the classical equation of motion is satisfied
[see Appendix A]. If the background field does not satisfy the equation of
motion one can integrate eq. (\ref{eom}) by parts which yields:
\be
S_{{\rm EoM}} = \frac{1}{12\pi^2}\!\int\!d^3 x\left\{E_i^a
     E_i^a\left(\frac{\pi}{|A_4|} - \frac{1}{T}\right) +
     \frac{\pi}{|A_4|^3}\left(E_i^a A_4^a\right)\left(E_i^b A_4^b\right) -
     \partial_i\left(E_i^a A_4^a\left(\frac{\pi}{|A_4|} -
         \frac{1}{T}\right)\right)\right\}.
\ee
Apart from the last term which is a full derivative the first two can be
added to the functions $f_{1,2}$ found previously. 

\section{Magnetic sector}

We are now going to calculate the fourth and fifth terms in eq. (\ref{proph}),
{\it i.e.} terms quadratic in the magnetic field. In analogy to the derivation of the action for the
electric sector, we make an expansion in the spatial covariant derivative $D_i$ of the
functional determinants and collect powers of the magnetic field. Note that while the
electric field only needed one power of a covariant derivative the magnetic field needs two.
For magnetic field squared we hence need an expansion to fourth order in the
covariant derivatives. In principle, in the fourth order in the covariant derivatives there is
a mixed term of the type $\epsilon_{ijk}\epsilon^{abc}E_i^aE_j^bB_k^c\,f(A_4)$, 
terms quartic in $E_i$ and terms containing covariant derivatives of $E_i$ but we do not
consider them here. For that reason,  we neglect all commutators $[D_i, A_4]$, as they
introduce additional powers of $E_i$. This simply means that we can drag all powers of the
covariant derivative $D_i$  as well as of the field strengths
$F_{ij}$  through the exponentials of $A_4$, as if they commute \footnote{One obtains from
the Jacobi identity $[F_{ij},A_4]=i\left([D_i,E_j]-[D_j,E_i]\right)$. Since we are
not interested now in such terms in the effective action, we shall assume that the magnetic
field commutes with $A_4$.}. 

For the ghost determinant (\ref{Dexp}) we obtain [see Appendix C.1] only one
gauge invariant structure, which after integration over $p$ becomes
\be\label{detDm}
\left[\log\,\det (-D^2)_{\rm{r,n}}\right]_{{\rm M}}^{(2)} =
\frac{1}{48\,\pi^{3/2}}\,\int\,d^3
x\,\sum_{k=-\infty}^{\infty}\int_0^{\infty}\frac{ds}{\sqrt{s}}\,\Tr\left(1-
e^{-s M^2}\right)\left(e^{s\A^2}\,B_k B_k \right)\;.
\ee

For the gluon determinant (\ref{Wexp}) we get the same result times a factor
of $-2$ plus one additional term:
\be
T_4\equiv\Tr \int_0^1\,d\alpha\,\int_0^{1-\alpha}\,d\beta\,e^{\alpha s \A^2} (2s\epsilon^{acb}F_{ij}^c)\,e^{\beta s
\A^2}(2s\epsilon^{dfe}F_{ij}^f)\,e^{(1-\alpha-\beta) s \A^2}\;, 
\ee
which with $F_{ij}^a F_{ij}^b = 2 B_k^a B_k^b$ and after integration
over $\alpha$ and $\beta$ yields the same gauge invariant structure:
\be
T_4 =  4\,s^2\Tr\left(e^{s\A^2} B_k B_k\right)\;. 
\ee
We find [see Appendix C.1] that this term yields the same contribution to the
action as the ghost determinant, but multiplied by a factor of $12$. The
total contribution of the gluon determinant to the 1-loop action in the
magnetic sector is hence $10$ times that of the
ghost determinant. 

The total 1-loop action in the magnetic sector is
\bea\label{Sm}
\lefteqn{\left[S_{\rm{1-loop}}^{(2)}\right]_{{\rm M}} = \left[\log\, \left(\det{W}\right)^{-1/2}_{{\rm
r,n}} +
  \log\,\det \left({-D^2}\right)_{{\rm r,n}}\right]^{(2)}_{{\rm M}}  }\\ \nn && =
\frac{11}{48\,\pi^{3/2}}\,\int\,d^3
x\,\sum_{k=-\infty}^{\infty}\int_0^{\infty}\frac{ds}{\sqrt{s}}\,\Tr\left(1-e^{-s
    M^2}\right)\left(e^{s\A^2}\,B_k B_k \right)\,,
\eea
where the integration over $s$ and the summation over Matsubara frequencies
still have to be performed. However, we do not need to do it anew since exactly the
same gauge invariant appeared in the invariant $I_2$ in the electric sector, 
with the obvious replacement $E_i\to B_i$, see \eq{I20}. With our gauge choice we obtain
two structures, $(B_i^1 B_i^1 + B_i^2B_i^2)\, h_1(\nu)$ and $B_i^3 B_i^3\, h_3(\nu)$, which
can be written in invariant terms as 
\be
B_i^2 h_1(\nu) \qquad{\rm and}\qquad \frac{(B_i\cdot A_4)^2}{|A_4|^2}
\left( h_3(\nu)-h_1(\nu)\right)\;,
\ee
respectively. 

Combining the ghost and the gluon result and denoting $(h_3-h_1)$ by $h_2$ we find the
two functions defined in eq.(\ref{proph}):
\bea\label{h1}
h_1(\nu) &=& \frac{f_3(\nu)}{2} +
\frac{11}{48\,\pi^2}\log\,\mu =  \frac{11}{96
\pi^2}\left[4\left(\log\,\frac{M}{4 \pi T} + \frac{\ge}{2}\right) - \psi\left(\nu\right) -
  \psi\left(1-\nu\right) \right] \,, \\
\label{h2}
h_2(\nu) &=& \frac{f_3(\nu)}{2} -\frac{11}{48\,\pi^2}\log\,\mu = -
\frac{11}{96 \pi^2}\left[2\ge + \psi\left(\nu\right) +
  \psi\left(1-\nu\right)\right]\,,\\
\label{h3}
h_3(\nu) &=& f_3(\nu) =  \frac{11}{48 \pi^2}\left[ 2 \log\,\frac{M}{4\pi T}  -
  \psi\left(\nu\right) - \psi\left(1-\nu\right)\right]\,, \qquad \nu=\frac{\sqrt{A_4^a A_4^a}}{(2\pi T)}\,.
\eea
As before $\mu$ is given by \eq{mu}. 

The UV divergent logarithm in \eq{h1} cancels the divergence from the tree-level action
which has the running coupling defined at the cutoff momentum $M$,
\be
\frac{B_i^a B_i^a}{2 g^2(M)} = B_i^a\,B_i^a\, \frac{11}{24 \pi^2}\,\log\frac{M}{\Lambda}\;,
\ee
such that the sum of tree and 1-loop actions are UV finite. The finite full action can be
brought into the form of eq.(\ref{effact}) where the functions $H_{1,2,3}$ are obtained from
$h_{1,2,3}$ by replacing the cutoff $M$ by $\Lambda$:
\bea
\label{H1}
H_1(\nu) &=& \frac{F_3(\nu)}{2} +
\frac{11}{48\,\pi^2}\log\,\mu = \frac{11}{96
\pi^2}\left[4\left(\log\,\frac{\Lambda}{4 \pi T} + \frac{\ge}{2}\right) - \psi\left(\nu\right) -
  \psi\left(1-\nu\right) \right] \;, \\
\label{H2}
H_2(\nu) &=&\frac{11}{96 \pi^2}\left[2\ge - \psi\left(\nu\right) - \psi\left(1-\nu\right)\right]\,,\\
\label{H3}
H_3(\nu) &=& F_3(\nu) =  \frac{11}{48 \pi^2}\left[ 2 \log\,\frac{\Lambda}{4\pi T}  
- \psi\left(\nu\right) - \psi\left(1-\nu\right)\right]\,,\qquad \nu=\frac{\sqrt{A_4^a A_4^a}}{(2\pi T)}\,.
\eea

When we sum over the Matsubara frequencies we find as in the case of the
electric sector, that the zero Matsubara frequency contributes both with a
finite term which is proportional to $1/\phi$ and a `naked' IR divergent
part. The finite terms are included in the results (\ref{h1})  and
(\ref{h2}). The separate contribution of the IR divergent part is 
\be\label{irsing}
\lim_{\ok\to 0}\frac{1}{\pi }\frac{11}{48}\int\,d^3 x\, \left(B_k^a  B_k^a - \frac{(B_i\cdot
A_4)^2}{|A_4|^2}\right)\,\frac{1}{\ok}\;.
\ee
In contrast to the electric sector, this divergence does not get canceled. This is also clearly
seen in terms of Feynman diagrams and corresponds to a singularity in the magnetic
self-energy of the gluons due to the zero Matsubara frequency in the loop, when the external
gluons have zero momentum. This singularity is regularized when higher terms in $B_i$
and/or nonzero momenta of the background magnetic field are taken into account.

\subsubsection{Magnetic sector without the zero Matsubara frequency contribution }

For a discussion of dimensional reduction at high temperatures we again remove the 
contribution of the zero Matsubara frequency and obtain:
\bea\label{H1t}
\tilde{H}_1(\nu) &=& \frac{11}{96 \pi^2}\left[4\left( \log\,\frac{\Lambda}{4 \pi T} + \frac{\ge}{2}\right)  -
\psi\left(\nu\right) - \psi\left(-\nu\right)\right] \;, \\
\label{H2t}
\tilde{H}_2(\nu) &=& -\frac{11}{96 \pi^2}\left[2 \ge + \psi\left(\nu\right) + \psi\left(-\nu\right)\right] \;, \\
\label{H3t}
\tilde{H}_3(\nu) &=& \frac{11}{48 \pi^2}\left[2 \log\,\frac{\Lambda}{4 \pi T}  -
\psi\left(\nu\right) - \psi\left(-\nu\right)\right]\;.
\eea
The IR singularity (\ref{irsing}) is of course not present.

\section{Comparison of our results to previous work}

There have been two other publications with the aim to get an effective
theory for QCD at high temperatures. In the first one \cite{Wirstam} gluon
by gluon scattering at low momenta in the nonzero temperature Yang-Mills
theory is calculated in terms of Feynman diagrams. In the second one
\cite{Chapman} an effective theory for the static modes of $SU(N)$
Yang--Mills theory in terms of a covariant derivative expansion is derived. The author makes
a series expansion of the functional determinants, and goes up to six orders in the covariant
derivative. This corresponds to a 1-loop calculation with up to six external static gluons and
their effective vertices. The zero Matsubara frequency contribution is not included. 

To compare our results to those of Refs.\cite{Wirstam} and \cite{Chapman} we
need to expand our functions $f_{1,2}$, $h_{1,2}$ in powers of $A_4$.  We
would like to stress that within our calculation we can go to arbitrary power
of $A_4$, while Refs. \cite{Wirstam} and \cite{Chapman}  only go to the quadratic order. For a
comparison we look at the quadratic terms in $A_4$ and obtain the following contribution:
\be
\int\,d^3 x\,\frac{\zeta(3)}{\pi^4 T^3}\frac{11}{384} \left(E^2 A_4^2 + 3 (E A_4)^2\right) \;.
\ee
This agrees precisely with Ref. \cite{Chapman} after collecting terms of the
orders above. Ref.\cite{Wirstam} differs both in sign and magnitude. For the magnetic part we
can only compare to Ref. \cite{Chapman}, as only there the terms under consideration have been
computed. To quadratic order in $A_4$ we obtain
\be
\int\,d^3 x\,\frac{\zeta(3)}{\pi^4 T^3}\frac{11}{192}\left(B^2 A_4^2 + (B A_4)^2\right)\;,
\ee
which again coincides exactly with the result derived in Ref. \cite{Chapman}. 
In addition, in Ref. \cite{BGK-AP} the 1-loop action for the Polyakov line has
been computed up to two derivatives in the specific case of zero $A_i$. 
Although it may look as being a gauge-noninvariant condition, in fact one of
the gauge-invariant structures can be extracted from that calculation. 
Indeed, the gauge-invariant combination $(E_iA_4)^2/A_4^2$ projects out
the $A_i$ field. Therefore, what we call the $f_3$ function has been actually 
computed in that paper, and our result coincides with theirs. 

\section{Time dependence and periodicity in $A_4$}

So far we have considered only time-independent background fields $A_4$ and $A_i$. 
As stressed in the Introduction, taking $A_4$ static is no restriction on the background but
merely a convenient gauge choice. However, taking $A_i$ static is a restriction, and we
would like to relax it, that is to include terms in the effective action containing time derivatives 
$\dot A_i$. To the second order in $\dot A_i$, this can be done in a very simple way.
Namely, we notice that in deriving the quantum action we have made use of the commutators
\urs{cDa}{cDaa} which remain exactly the same if we replace $A_4$ in $\A$ by the more general
operator $D_4$. The only difference is that the resulting electric field should be now
understood as the full $E_i^a=D_i^{ab}A_4^b-\dot A_i^a$. With this replacement, one gets
the same effective action \ur{effact} as in the case of a purely static $A_i$. As in the static
case, it is limited to the second power of $E_i$ and hence of $\dot A_i$. Therefore, its
applicability is restricted by the condition that both spatial and time derivatives of the fields
are much less than the temperature.   

After fixing the gauge such that $A_4$ is static one can perform further a time-independent
gauge rotation to make $A_4(x)$ diagonal {\it i.e.} belonging to the Cartan subalgebra
at all spatial points. We shall use this gauge condition in this section to simplify the
discussion. For the $SU(2)$ gauge group it means that we take $A_4=\phi\,\frac{\tau^3}{2}$.

Having fixed the gauge such that $A_4$ is static and diagonal there is only an Abelian
residual gauge symmetry left. It consists of arbitrary time-independent gauge rotations about
the Cartan generators, and of a timedependent gauge rotation (also about the Cartan axes)
of a special discrete type compatible with periodicity of $A_i(x,t)$. For the $SU(2)$ gauge
group this residual gauge symmetry is with respect to the Abelian gauge transformation 
\be\label{Ab}
A_\mu\to S^\dagger A_\mu S+iS^\dagger\partial_\mu S,\qquad 
S(x,t)=\exp\left\{-i\frac{\tau^3}{2}\left[\alpha(x)+2\pi tTn\right]\right\},
\ee
where $n$ is an integer, which follows from the requirement that $A_i(x,t)$ remains periodic
in time. One cannot take rotations about axes other than the $3^{\rm d}$ one because it will
make $A_4$ nondiagonal, and one cannot take the time dependence other than linear
because that would make $A_4$ timedependent. In components, the transformation \ur{Ab}
reads:
\bea
\label{A43}
A_4^{\prime\,3}(x)&=&A_4^3(x)+2\pi T\, n,\qquad {\rm meaning}\quad\nu^\prime=\nu+n,\\
\label{Ai1}
A_i^{\prime\,1}(x,t)&=&\cos\beta A_i^1+\sin\beta A_i^2,\qquad 
\beta(x,t)=\alpha(x)+2\pi T n t,\\
\label{Ai2}
A_i^{\prime\,2}(x,t)&=&-\sin\beta A_i^1+\cos\beta A_i^2, \\
\label{Ai3}
A_i^{\prime\,3}(x,t)&=&A_i^3+\partial_i\alpha(x).
\eea
The effective action must be invariant under this transformation, but is it?

It is easy to check that the combinations of the field strengths $B_i^3B_i^3$, $B_i^\perp
B_i^\perp$, $E_i^3E_i^3$ and $E_i^\perp E_i^\perp$ (where $F^\perp F^\perp$ is the
short-hand notation for $F^1F^1+F^2F^2$) are invariant under the gauge transformation
(\ref{A43}-\ref{Ai3}). As follows from \eq{effact} these structures are multiplied by the
functions $H_3(\nu),H_1(\nu),F_3(\nu)$ and $F_1(\nu)$, respectively. Therefore to support
the invariance of the effective action under the gauge transformation \ur{Ab}, all the four
functions need to be {\em periodic} in $\nu=A_4^3/2\pi T$ with unit period. 

So far we have computed those functions in the domain $0<\nu<1$, so to check the
periodicity one has to know them outside this domain. Actually only the last step, namely
the summation over Matsubara frequencies, has to be revisited. The result is as follows:
The static potential $V(\nu)$ and the functions $H_1(\nu),H_3(\nu)$ and $F_3(\nu)$ are,
indeed, periodic (and even) in $\nu$, whereas the last function $F_1(\nu)$ is even but 
{\em not} periodic.    

Indeed, for $0<|\nu|<1$ we find:
\be
F_1(\nu)= -\frac{11}{48\pi^2}\left[L + \po\left(\frac{\nu}{2}\right)+
  \po\left(-\frac{\nu}{2}\right)-\frac{20}{11|\nu|}\right]\;.
\ee
For $1<|\nu|<2$ we find:
\be
F_1(\nu) =  -\frac{11}{48\pi^2}\left[L + \po\left(\frac{\nu}{2}\right)+
  \po\left(-\frac{\nu}{2}\right)-2-\frac{38}{11\nu}
  -\frac{8}{11}\left(\frac{1}{\nu^2} - \frac{1}{|\nu^3|}\right)\right]\;.
\ee
For $2<|\nu|<3$ we find:
\be
F_1(\nu) =  -\frac{11}{48\pi^2}\left[L  + \po\left(\frac{\nu}{2}\right)+
  \po\left(-\frac{\nu}{2}\right)-3-\frac{56}{11|\nu|}
  -\frac{24}{11\nu^2} -\frac{40}{11|\nu^3|}+\frac{2}{2-|\nu|}+\frac{2}{4-|\nu|}\right]\,,
\ee
etc. By $L$ we have denoted the constant part: $L=-2 \log(\Lambda/4\pi T)$.This function
is plotted in Fig. \ref{F1fig} and is clearly not
periodic.

\begin{figure}[t]
\centerline{
\epsfxsize=0.45\textwidth
\epsfbox{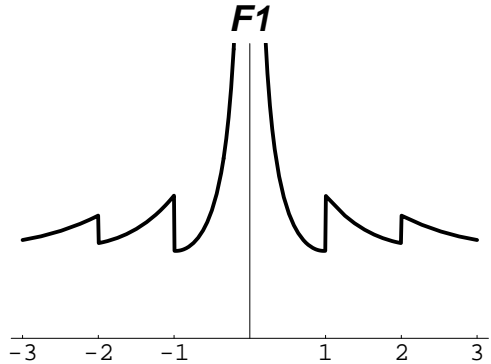}}
\caption{ {\small The function $F_1(\nu)$ with the constant part $L$ subtracted, in different
intervals.}\label{F1fig} }
\end{figure}

The reason of this periodicity paradox is clear. By making the timedependent gauge
transformation (\ref{A43}-\ref{Ai3}) we induce large time derivatives of the $A_i^{1,2}$
fields, being of the order of $2\pi T$. The `dynamical' electric field $\dot A_i^{1,2}$ does
not enter into the invariants $B_i^3B_i^3$, $B_i^\perp B_i^\perp$ and $E_i^3E_i^3$.
Therefore, the corresponding functions $H_1(\nu),H_3(\nu)$ and $F_3(\nu)$ should
be periodic to support gauge invariance, which indeed they are. One cannot and should
not observe gauge invariance in the structure $E_i^\perp E_i^\perp\,F_1(\nu)$ as it is only 
a quadratic functional in $\dot A_i^{1,2}$, which is insufficient. There is no periodicity
requirement on $F_1(\nu)$. All powers of  $\dot A_i^{1,2}/T$ (and of $A_4/T$) need to be
collected in the effective action to check the invariance under fast timedependent gauge
transformations \ur{Ab}. This, however, lies beyond the scope of the present study.  


\section{Conclusions}


Given that the Polyakov line in the $A_4$ static gauge is 
\be\label{Pol}
P(x)=\exp\left(i\frac{A_4(x)}{T}\right),
\ee
its gauge-invariant eigenvalues are $e^{\pm i\pi\nu}$ where $\nu=\sqrt{A_4^aA_4^a}/2\pi T$.
We have in fact computed the 1-loop effective action for the eigenvalues of the Polyakov 
line, interacting with the spatial components of the Yang--Mills field $A_i$. For the $SU(2)$
gauge group the effective action is given by \eq{effact} with the four functions
$F_{1,2}(\nu)$ and $H_{1,2}(\nu)$ defined in \eqs{F1}{F2} and \eqs{H1}{H2}, respectively.
All functions are singular and behave as $1/A_4$ at small $A_4$, which is due
to the contribution of the zero Matsubara frequency. It should be stressed that the
functions $H_{1,2}(\nu)$ being coefficients in front of the invariants quadratic in the magnetic
field, contain `naked' IR divergences which are regularized by higher orders of 
the magnetic field and/or field momentum. If the static fluctuation mode is excluded from the
effective action, all functions become finite and nonsingular; they are then given by
\eqs{F1t}{F2t} and \eqs{H1t}{H2t}, respectively. If the background field does not satisfy
the Yang--Mills equations of motion, there is an additional term \ur{eom}.

As it should be expected, the functions $H_{1,2}(\nu),\,F_3(\nu)=F_1(\nu)+F_2(\nu)$ and the
static potential $V(\nu)$ are periodic functions of $\nu$ but $F_1(\nu)$ is not. The periodicity
is related to the gauge invariance of the effective action with respect to fast timedependent
gauge transformations inducing large electric field $\dot A_i$. For the particular structure
related to $F_1(\nu)$, this gauge invariance can only be revealed when all powers of the
electric field in the effective action are collected. 

For higher gauge groups there will be more invariant structures already in the quadratic
order in the electric and magnetic fields, and the coefficient functions will depend
on all the eigenvalues of the Polyakov line, whose number is $N-1$ for the $SU(N)$
gauge group. It is worthwhile to generalize this work to higher groups, as well as to
find the 1-loop affective action arising from integrating out fermions. 

One can think of two kinds of applications of our results. One is for studying correlation
functions, say, of the Polyakov lines at high temperatures but going beyond
the approximations used previously. One might be also interested in evaluating the
1-loop weights of extended semiclassical objects, such as calorons with nontrivial
holonomy and dyons. The technique developed in this paper is applicable for
such studies.   

\noi
{\sc Acknowledgments:}\\
We are grateful to Rob Pisarski for valuable discussions and to Eric Braaten for reading the
manuscript and helpful comments. D.D. would like to thank Victor Petrov and M.O. would like
to thank Jonathan Lenaghan for useful conversations at the early stage of this work.

M.O. thanks Brookhaven National Laboratory for its hospitality and was partly supported by the U.S. Department of Energy under Contract No. DE-AC02-98CH10886. 

\begin{appendix}
\section{Notations}

We normalize the generators of an SU(N) group as 
\be
\Tr\,t^a\,t^b = \frac{1}{2}\,\delta^{ab}\;.
\ee
For SU(2) $t^a$ are half the Pauli matrices, and for SU(3) half the Gell-Mann
matrices. Their commutator defines the generators of the adjoint representation:
\be
(T_a)^{bc} = -i\,f^{abc} = i\,f^{acb}\;.
\ee
The field strength in the fundamental representation is defined as
\be
[\nabla_\mu, \nabla_\nu] = - i\, F_{\mu\nu} = -i\, F_{\mu\nu}^a\,t^a\;,
\ee
where 
\be
\nabla_{\mu} = \partial_{\mu} - i\,A_{\mu}^a\,t^a\;,
\ee
is the covariant derivative in the fundamental representation. 

The field strength in adjoint representation becomes
\be
[D_\mu, D_\nu]^{cd} = - f^{cda}\,F_{\mu\nu}^a = f^{cad}\,F_{\mu\nu}^a\;,
\ee
where $D_{\mu}$ is the covariant derivative in the adjoint representation:
\be
D_{\mu}^{ab} = \partial_{\mu}\delta^{ab} + f^{acb}\,A_{\mu}^c\;.
\ee
For any matrix in the adjoint representation we shall imply $B^{ab} = -
i\,f^{abc}\,B^c$. In particular, the gauge field which in the fundamental
representation is $A_{\mu} = A_{\mu}^a\,t^a$ becomes in the adjoint representation
\be
A_{\mu}^{cd} = -i\,f^{cde}\,A_{\mu}^{e} = i\,f^{ced}\,A_{\mu}^{e}\;.
\ee
The electric field is in general defined as $E_i\equiv F_{i4}$. Hence 
\be
[D_i, D_4] = -i\,F_{i4} = -i\,E_i\;.
\ee
Explicitly in the adjoint representation one has
\be\label{ae}
E_i^{ab} = -i\,f^{abd}D_i^{dh}A_4^h =  -i\,f^{abd} E_i^d \qquad{ \rm and}\qquad
E_i^d = D_i^{dh}A_4^h\;.
\ee
We notice that the combination
\be\label{ade}
[D_i, E_i]^{ab} = -i f^{abe}\,D_i^{ec}\,E_i^c\;
\ee
is zero if the background field satisfies the Yang--Mills equation of motion,
$D_\mu^{ac} F_{\mu\nu}^{c} = 0$. 
For eq. (\ref{ade}) we use the Jacobi identity
\be
f^{abe}f^{efc} = f^{afe}f^{ebc} - f^{ace}f^{ebf}\;.
\ee

\section{Functional determinants in the electric sector}

\subsection{Managing functional traces}

We are interested here in extracting terms quadratic in the electric field
but having any power of $A_4$. We expand the ghost functional determinant
(\ref{Dexp}) to quadratic order in $D_i$ with the help of eq. (\ref{master1})
and obtain two contributions at this order:
\be\label{t1}
T_1 = \Tr\int_0^1\,d\alpha\,e^{\alpha s \A^2} sD_i^2 \,e^{(1-\alpha) s \A^2}\;,
\ee
and
\be\label{t2}
T_2 = -\frac{4}{3}s^2 p^2\Tr\int_0^1 d\alpha \int_0^{1-\alpha}d\beta\,e^{\alpha s \A^2} D_i\,e^{\beta s \A^2}
D_i\,e^{(1-\alpha -\beta) s \A^2}\;,
\ee
where for $T_2$ we used the fact that $p_i p_j$ averaged over the directions
of the
three-vector $p_i$ gives $\frac{1}{3} \delta_{ij} p^2$. With the commutators (\ref{master2}) and
(\ref{cDa}) it can be shown that $T_1$ is a sum of four terms, two of which are gauge
invariant and
two are not gauge invariant (denoted by a bar):
\be
T_1 = T_{11} + T_{12} + \bar{T}_{11}+ \bar{T}_{12}\;,
\ee
where
\bea\label{t1terms}
T_{11} &=& -s^2\int_0^1 d\alpha\,(1-\alpha)\Tr\,e^{s\A^2}\left(2 E_i^2 + i\left\{\A, \left[D_i,
E_i\right]\right\}\right)\;, \\
T_{12} &=& -2s^3\int_0^1 d\alpha \int_0^1 d\gamma \int_0^1 d\delta \,(1-\alpha)^2 (1-\gamma) \Tr
\,e^{s\A^2[1-\delta(1-\gamma)(1-\alpha)]}\left\{\A,E_i\right\} e^{s\A^2 \delta (1-\gamma)(1-\alpha)} \left\{\A,E_i\right\}\;\nn
\\ 
 \bar{T}_{11} &=& s \,\Tr\left(e^{s\A^2}D^2\right)\;,\nn \\ 
\bar{T}_{12} &=& -2is^2\int_0^1 d\alpha \int_0^1 d\gamma \,(1-\alpha)\Tr \,e^{s\A^2[\alpha +
\gamma(1-\alpha)]}\left\{\A,E_i\right\}e^{s\A^2 (1-\gamma)(1-\alpha)} D_i \nn \;.
\eea
As the action is gauge invariant, we expect the not gauge invariant terms to cancel with
those of the term $T_2$. We will show, that this is indeed the case. 

Let us now turn to $T_2$. Dragging $D_i$'s to the right we obtain:
\bea
T_2 = &-&\frac{4}{3}s^2 p^2\int_0^1 d\alpha \int_0^{1-\alpha}d\beta\,e^{(1-\alpha) s \A^2}D^2\,e^{\alpha s \A^2}
\\  
&+& \frac{4i}{3}s^3 p^2\int_0^1 d\alpha \int_0^{1-\alpha}d\beta\,\beta\int_0^1 d\gamma\, \Tr\, e^{(1-\alpha -\beta +\gamma\beta) s
\A^2}\left\{\A,E_i\right\}e^{(1-\gamma)\beta s\A^2}\\ 
\nn
&\times & \left\{e^{\alpha s \A^2}D_i - is\alpha \int_0^1 d\gamma\,e^{\delta\alpha s\A^2}\left\{\A,E_i\right\} 
e^{(1-\delta)\alpha s\A^2}\right\}\;.
\eea
We find that $T_2$ is a sum of six terms, amongst which three are gauge invariant and three
(again denoted by a bar) are not. We start with the not gauge invariant ones and show that
they cancel with $\bar{T}_{11}$ and $\bar{T}_{12}$
\bea
\bar{T}_{21} &=& -\frac{2}{3} s^2 p^2 \Tr\,\left(e^{s\A^2}D^2\right),\\
\bar{T}_{22a} &=& \frac{8i}{3} s^3 p^2\int_0^1 d\alpha\, \alpha(1-\alpha) \int_0^1 d\gamma\,
\Tr\,e^{s\A^2[\alpha+\gamma(1-\alpha)]} \left\{\A,E_i\right\}e^{s\A^2(1-\gamma)(1-\alpha)}\,D_i,\\
\bar{T}_{22b} &=& \frac{4i}{3} s^3 p^2\int_0^1 d\alpha \int_0^{1-\alpha}d\beta\,\beta\int_0^1
d\gamma\,\Tr\,e^{s\A^2(1-\alpha-\beta+\gamma\beta)} \left\{\A,E_i\right\}\,e^{s\A^2[\alpha+\beta(1-\gamma)]}\,D_i\,.
\eea
The terms $\bar{T}_{11}$ from eq. (\ref{t1terms}) and $ \bar{T}_{21}$ are of the same structure.
Their sum is
\be
\bar{T}_{11}+ \bar{T}_{21} =  s\left[1-\frac{2 s p^2}{3}\right]\, \left(e^{s\A^2}D^2\right)\;.
\ee
This term vanishes upon $p$ integration since:
\be
\int_0^{\infty}d^3 p\,sp^2\,e^{-sp^2} = \frac{3}{2}\int_0^{\infty}d^3 p\,e^{-sp^2}\;.
\ee
For the evaluation of the other non gauge-invariant terms we use some relations for
integrations over parameters, valid for any function $f$:
\bea
\int_0^1 d\alpha \int_0^1 d\gamma \,\alpha(1-\alpha)\,f(\underbrace{\alpha\gamma}_{\epsilon}) &=& \int_0^1
d\epsilon\,\frac{(1-\epsilon)^2}{2} f(\epsilon)\;,\\
\int_0^1 d\alpha \int_0^1 d\gamma \,\alpha\,f(\underbrace{\alpha\gamma}_{\eta}) &=& \int_0^1
d\eta\,(1-\eta)\,f(\eta)\;,\\
\int_0^1 d\alpha \int_0^{1-\alpha} d\beta \beta \int_0^1 d\gamma\,f(\underbrace{\alpha + \beta\gamma}_{\delta}) &=& \int_0^1 d\delta\,
\delta(1-\delta)\, f(\delta)\;.
\eea
We find that
\be
\!\!\!\bar{T}_{12} \!+ \!\bar{T}_{22a} \!+ \!\bar{T}_{22b} \!\!=  \!\!
\int_0^1 \!\! d\alpha \!\! \left[ -2is^2 (1\!-\!\alpha) \!\! + \!\! \frac{4 i}{3}s^3 p^2
  (1\!-\!\alpha)^2 \!\! + \!\! \frac{4 i}{3}s^3 p^2 \alpha(1\!-\!\alpha)\right] \!
\Tr\, e^{s\A^2(1\!-\!\alpha)} \!\! \left\{\A, E_i\right\} \! e^{s\A^2 \alpha}D_i \;.
\ee
After integration over $\alpha$ the term in the brackets
\be
\left[\ldots\right] = -is^2\left(1 - \frac{2}{3}sp^2\right)
\ee
becomes zero after integration over $p$. We hence have shown that all not gauge invariant
terms cancel out, as was indeed expected. 

There are three gauge invariant terms left. They can be simplified by using some more
integration relations:
\bea
\int_0^1 d\alpha\,\alpha^2 (1-\alpha) \int_0^1d\gamma\,\gamma \int_0^1 d\delta\,f(\underbrace{\alpha\gamma\delta}_{\xi}) &=&
\frac{1}{6}\int_0^1 d\xi\,(1-\xi)^3\,f(\xi) \;,\\
\int_0^1 d\alpha\,\alpha \int_0^{1-\alpha}d\beta\,\beta\int_0^1d\gamma\int_0^1 d\delta\, f(\underbrace{\alpha\delta+\beta\gamma}_{\zeta})
&=& \int_0^1\,\frac{\zeta(1-\zeta)^2}{2}\,f(\zeta)\;,\\
\int_0^1 d\alpha\,\alpha^2 \int_0^1 d\gamma\,\gamma \int_0^1 d\delta \, f(\underbrace{\alpha\gamma\delta}_{\kappa}) &=&
\frac{1}{2}\int_0^1 d\kappa\,(1-\kappa)^2\,f(\kappa)\;.
\eea
Using them we find two gauge invariant structures
\bea\label{I1}
I_1 &\equiv& T_{12}+T_{22a}+T_{22b} \\ &=& s^3\int_0^1
d\alpha\,\left\{-\frac{1}{2} + \alpha(1-\alpha)+\frac{2}{9}s
  p^2\left[1-\frac{3}{2}\alpha(1-\alpha)\right]\right\}\,\Tr\,e^{(1-\alpha)
  s\A^2}\left\{\A, E_i\right\}\, e^{\alpha s\A^2}\,\left\{\A, E_i\right\} \nn \;,\\ 
I_2&\equiv& T_{11}+T_{21} = -s^2\left( \frac{1}{2}-\frac{2}{9}sp^2 \right)\, \Tr\,e^{s\A^2}\left(2
E_i^2 + i\left\{\A,\left[D_i, E_i\right]\right\}\right)\label{I2}\;.
\eea

In the evaluation of the gluon determinant (\ref{Wexp}) there is one more
gauge invariant structure: 
\bea\label{I3}
I_3 \equiv  T_3  &=&  -2\int_0^1\, d\alpha \int_0^{1-\alpha}\, d\beta \, \Tr\, e^{\alpha s\A^2} \,
(2s E_i)\, e^{\beta s\A^2}(2s E_i)\, e^{(1-\alpha-\beta)s\A^2} \\ 
 &=&  -4 s^2\int_0^1\, d\alpha\, \Tr \,e^{(1-\alpha) s\A^2} E_i\, e^{\alpha s\A^2} E_i\;.
\eea

\subsection{Integrating over $\alpha$, $s$, $p$ and summing over Matsubara frequencies}

To obtain the action we have to integrate over $\alpha$, $s$, momentum $p$ and sum over
Matsubara frequencies $\ok$ for the three invariants we derived. For convenience we take
out a factor of $1/(2\pi^2T)$:
\bea
\frac{1}{2\pi^2 T}\int\,d^3 x\,\sum_{k=-\infty}^{\infty}\int\frac{d^3 p}{(2
\pi)^3}\,\int_0^{\infty}\frac{ds}{s}\,e^{-s p^2}\,2\pi^2T\,I_j
\qquad{{\rm where}}\qquad
\qquad{j=1,2,3}\qquad
\;.
\eea

\subsubsection{The first invariant $I_1$}

After taking the trace in eq. (\ref{I1}) explicitly and integrating over $\alpha$ we
find that $I_1$ has the structure:
\be
2\pi^2T\,I_1 = X_1\,(E_i^1\,E_i^1 + E_i^2\,E_i^2) + X_3\,E_i^3\,E_i^3\;.
\ee
This is expected since our gauge choice for $A_4$ is along the
third color direction, so the result should by symmetric in $E^{1,2}$. Next we
integrate over momentum, using that $d^3 p = 4\pi\,p^2\,dp$. We find
\bea
2 \pi^2 P_1 &\equiv& \int_0^{\infty}dp\,p^2\, X_1 = p_{10}\,e^{-s \ok^2} + p_{1p}\,e^{-s (\phi +\ok)^2} +
p_{1m}\,e^{-s (\phi -\ok)^2}\;,\\
2 \pi^2 P_3 &\equiv& \int_0^{\infty}dp\,p^2\, X_3 = p_{3p}\,e^{-s (\phi +\ok)^2} + p_{3m}\,e^{-s (\phi
-\ok)^2}\;,
\eea
where the coefficients are:
\bea
p_{10} &=& \frac{\sqrt{\pi}\,(-3 + s\phi^2)}{12\, s^{3/2}\,\phi^2} -
\frac{\sqrt{\pi}}{4\,s^{5/2}\,\phi^3\,(-\phi + 2\ok)} + \frac{\sqrt{\pi}}{4\,s^{5/2}\,\phi^3\,(\phi + 2\ok)} \;,\\ 
p_{1p} &=& -\frac{\sqrt{\pi}\,(3 + s\phi^2)}{24\, s^{3/2}\,\phi^2} - \frac{\sqrt{\pi}\,\ok}{12\, \sqrt{s}\,
\phi} - \frac{\sqrt{\pi}}{4\,s^{5/2}\,\phi^3\,(\phi + 2\ok)} \;,\\
p_{1m} &=& -\frac{\sqrt{\pi}\,(3 + s\phi^2)}{24\, s^{3/2}\,\phi^2} + \frac{\sqrt{\pi}\,\ok}{12\, \sqrt{s}\,
\phi} + \frac{\sqrt{\pi}}{4\,s^{5/2}\,\phi^3\,(-\phi + 2\ok)}\;,\\
p_{3p} &=& \frac{\sqrt{s\,\pi}\,(\phi +\ok)^2}{12}\;,\\
p_{3m} &=& \frac{\sqrt{s\,\pi}\,(\phi -\ok)^2}{12}\;.
\eea
Next we integrate over $s$. The integrals over $s$ of the individual terms in
$P_1$ turn out to be UV divergent. however their sum is finite. 
Using a regularization, {\it i.e.} replacing the integration kernel $P_1$ by
$\lim_{\epsilon\to 0}\left(P_1\cdot s^{\epsilon}\right)$ we find the finite result:
\bea\label{S1}
S_1 &\equiv& \lim_{\epsilon\to 0}\int_0^{\infty}ds\,\left(P_1\cdot s^{\epsilon}\right) = \frac{\pi\,|\ok|\,(\phi^4
+ 2\, \phi^2\,\ok^2 - 16\,\ok^2)}{12\,\phi^2\,\ok^2\,(\phi^2 - 4\,\ok^2)} -\frac{\pi\,(\phi +
2\,\ok)}{24\,\phi\,|\phi + \ok|}\\ \nn
&+& \frac{\pi\,|\phi + \ok|}{4\,\phi^2} -  \frac{\pi\,|\phi + \ok|^3}{3\,\phi^3\,(\phi + 2\,\ok)} 
 -\frac{\pi\,(\phi - 2\,\ok)}{24\,\phi\,|\phi - \ok|} + \frac{\pi\,|\phi - \ok|}{4\,\phi^2} -  \frac{\pi\,|\phi -
\ok|^3}{3\,\phi^3\,(\phi - 2\,\ok)}\;.
\eea
The integral over $P_3$ is finite and we obtain
\be\label{S3}
S_3\equiv\int_0^{\infty}ds\,P_3 = \frac{\pi\,(\phi - \ok)^2}{24\,|\phi - \ok|^3} + \frac{\pi\,(\phi +
\ok)^2}{24\,|\phi + \ok|^3}\;.
\ee
The next and final step is to sum over the Matsubara frequencies. We replace
$\ok$ once by $2\pi T k$ and once by $-2\pi T k$, where $k\geq 1$, and add
the two results, which has the advantage, that we have to sum over the
positive frequencies only. The zero Matsubara frequency will be treated
separately. The field variable is rescaled according to
\be
\phi = 2 \pi T \nu 
\qquad {{\rm where}} \qquad
0\le \nu \le 1 \;.
\ee
This results into 
\bea\label{S1'}
T S_1' &=& \frac{1}{12k} + \frac{1}{12(\nu-2k)} - \frac{1}{24(\nu-k)} + \frac{1}{24(\nu+k)} -
\frac{1}{12(\nu+2k)}\\\label{S3'}
T S_3' &=& \frac{1}{24(\nu+k)} -\frac{1}{24(\nu-k)}\;,
\eea
where the prime indicates, that this is valid for nonzero Matsubara
frequencies. 

The contribution of the zero Matsubara frequency in eq. (\ref{S1}) consists of a
finite and a divergent part. The former becomes upon rescaling
\be\label{l10}
l_1^{(0)}(\nu) = T \left(\lim_{\ok\to 0} S_1\right)_{{\rm finite}} =  -\frac{1}{8\,\nu}\;.
\ee
We shall show later that the divergent part  cancels exactly with a divergent term from the
second invariant $I_2$. The sum over the first term in eq. (\ref{S1'}) is logarithmically
divergent. We regularize it by introducing a cutoff $\mu$ in the sum. This is
equivalent to the Pauli-Villars regularization, since we find that the
cutoff $\mu$ is related to the Pauli-Villars mass by
\be
\mu=\frac{M}{4\pi T}\,e^{\gamma_E}
\ee
where $\gamma_E$ is Euler's constant. 
Summing over the nonzero Matsubara frequencies and adding the finite contribution
from the zero mode (\ref{l10})we find
\be\label{l1}
 \! \!l_1(\nu) \!= \! T \left(\sum_{k=1}^{\infty}S_1'\right) \! + \! l_1(\nu)^{(0)} \! = \!
 \frac{1}{24}\left[ \! - \frac{3}{\nu} \! - \! \po\left(1 \!- \!\nu\right) \!
   - \! \po\left(1 \!+ \!\nu\right) \! + \! \po\left( 1 \! \!- \!\frac{\nu}{2}
     \!\right) \! + \! \po\left( \! 1 \!+ \!\frac{\nu}{2} \!\right) \! + \!
   2\,\log\,\mu \right]
\ee
where the $\psi$-function is the logarithmic derivative of the gamma function:
\be
\psi(z) = \frac{\partial}{\partial\,z}\log\,\Gamma(z)\;.
\ee

In the case of $S_3$ the zero Matsubara frequency yields only a finite part:
\be\label{l30}
l_3^{(0)}(\nu) = T \lim_{\ok\to 0}S_3 =  \frac{1}{24\,\nu}\;.
\ee
For the remaining sum over $S_3'$, we add and subtract $1/k$ terms to make them
convergent:
\be
\sum_{k=1}^{\infty}\left(\frac{1}{24(\nu+k)} - \frac{1}{24\,k}\right) -\left(\frac{1}{24(\nu-k)} +
\frac{1}{24\,k}\right) + \frac{1}{12\,k}  \;.
\ee
The first two terms yield $\psi$ functions, and the last part becomes a
logarithm after we introduce the cutoff $\mu$ in the sum over $1/k$. The sum
over all frequencies finally yields
\be\label{l3}
l_3(\nu) = T \left(\sum_{k=1}^{\infty}S_3'\right) + l_3^{(0)}(\nu) = \frac{1}{24}\left[ -2 \ge -
\po\left(\nu\right) - \po\left(1-\nu\right) + 2\,\log\,\mu \right]\;.
\ee

\subsubsection{The second invariant $I_2$}

We take the trace in eq. (\ref{I2}) but without the term 
$\left\{\A,\left[D_i,E_i\right]\right\}$  which is zero if the equation of motion is satisfied
by the background field (\ref{eom}). Integration over $\alpha$ we find the structure:
\be
2\pi^2T\,I_2 = Y_1\,(E_i^1\,E_i^1 + E_i^2\,E_i^2) + Y_3\,E_i^3\,E_i^3\;.
\ee
Next we integrate over momentum and obtain:
\bea\label{Q1}
2 \pi^2 Q_1 &\equiv& \int_0^{\infty}dp\,p^2\, Y_1 = -\frac{\sqrt{\pi}}{24\,\sqrt{s}}\left(2\,e^{-s \ok^2} +
e^{-s (\phi +\ok)^2}  + e^{-s (\phi -\ok)^2}\right)\\ \label{Q3}
2 \pi^2 Q_3 &\equiv& \int_0^{\infty}dp\,p^2\, Y_3 = -\frac{\sqrt{\pi}}{12\,\sqrt{s}}\left(e^{-s (\phi +\ok)^2} 
+ e^{-s (\phi -\ok)^2}\right)\;.
\eea
We integrate these functions over s and find:
\bea\label{R1}
R_1 &\equiv& \int_0^{\infty}ds\,Q_1 = -\frac{\pi}{24}\left(\frac{1}{|\phi + \ok |} + \frac{1}{|\phi -
\ok|} + \frac{2}{|\ok|}\right) \\ \label{R3}
R_3 &\equiv&\int_0^{\infty}ds\,Q_3 = -\frac{\pi}{12}\left( \frac{1}{|\phi - \ok|} + \frac{1}{|\phi +
\ok|}\right) \;.
\eea
For the sum over the nonzero Matsubara frequencies we replace $\ok$ once by
$2\pi T k$, once  by $-2\pi T k$ and add the two results. This yields:
\bea\label{R1'}
T R_1' &=& -\frac{1}{12k} + \frac{1}{24(\nu-k)} - \frac{1}{24(\nu+k)} \\\label{R3'}
T R_3' &=& \frac{1}{12(\nu-k)} -\frac{1}{12(\nu+k)}\;.
\eea
The contribution of the zero Matsubara frequency in $R_1$ yields a finite and
a divergent part where the former is
\be\label{g10}
g_1^{(0)}(\nu) = T \left(\lim_{\ok\to 0}R_1\right)_{{\rm finite}} =  -\frac{1}{24\,\nu}\;.
\ee
The `naked' $1/\ok$ divergencies came from (\ref{S1}) in $I_1$ and from
(\ref{R1}) in $I_2$. Both in the ghost and gluon determinants the two invariants
enter in the combination $I_1+I_2$. Adding the terms which produce the divergences 
and expanding in $\ok$ we find
\be
\frac{\pi\,|\ok|\,(\phi^4 + 2\, \phi^2\,\ok^2 - 16\,\ok^2)}{12\,\phi^2\,\ok^2\,(\phi^2 - 4\,\ok^2)} -
\frac{\pi}{12\,|\ok|} = \frac{\pi\ok}{2\,\phi^2} + {\cal{O}}(\ok^3)\;,
\ee
which obviously disappears as $\ok\to 0$. Therefore the divergent parts from the two
invariants cancel each other. 

Setting $\ok = 0$ in $R_3$ gives only a finite result:
\be\label{g30}
g_3^{(0)}(\nu) = T \left(\lim_{\ok\to 0}R_3 \right)=  -\frac{1}{12\,\nu}\;.
\ee
To sum over the nonzero Matsubara frequencies we use the
method of adding and subtracting $1/k$ terms to make individual sums
convergent and introduce the cutoff $\mu$ for divergent sums over
$1/k$ terms. Adding all contributions we obtain:
\bea\label{g1}
g_1(\nu) &=& T \left(\sum_{k=1}^{\infty}R_1'\right) + g_1(\nu)^{(0)} = 
 \frac{1}{24}\left[2\,\ge + \po\left(\nu\right) +
   \po\left(1-\nu\right)  - 4\,\log\,\mu \right]\;, \\ 
g_3(\nu) &=& T \left(\sum_{k=1}^{\infty}R_3'\right) + g_3(\nu)^{(0)} =
\frac{1}{12}\left[2\,\ge + \po\left( \nu\right) + \po\left(1-\nu\right)  - 2\,\log\,\mu
\right]\label{g3}\;.
\eea

\subsubsection{The third invariant $I_3$}

As in the case of the first two invariants we integrate and sum over the
third one (\ref{I3}). We find again the same structure in the electric
fields. Integrations over $s$ turn out to be finite. In the sum over
$\ok$ the zero mode only yields a finite part, and for the nonzero modes we
introduce the  cutoff $\mu$ in the sum over $1/k$ terms.  The calculation is
similar to the previous case, and we present here only the results:
\bea\label{j1}
j_1(\nu) &=& \left[ \frac{2}{\nu} - 2\,\ge - \po\left(-\frac{\nu}{2}\right) - \po\left(\frac{\nu}{2}\right) 
+ 2\,\log\,\mu \right]\;,\\
j_3(\nu) &=& \left[- 2\,\ge - \po\left(1-\nu\right) - \po\left(\nu\right) +
2\,\log\,\mu \right]\label{j3}\;,
\eea
where $j_1$ is the function in front of $(E_i^1 E_i^1 + E_i^2 E_i^2)$ and
$j_3$ multiplies $E_i^3 E_i^3$. 

\section{Functional determinants in the magnetic sector}

\subsection{Managing functional traces}

We are looking for terms quadratic in the magnetic field
but containing any power of $A_4$. Since we are not interested in terms
containing the electric field, we can drag all powers of covariant derivatives through the
exponentials of $A_4$ as if they commute. For the ghost contribution this gives
\bea
\lefteqn{\left[\log\,\det (-D^2)_{\rm{n}}\right]_{{\rm M, \, ghost}}^{(2)}=  - \int\,d^3
x\,\sum_{k=-\infty}^{\infty}\int\frac{d^3 p}{(2
\pi)^3}\,\int_0^{\infty}\frac{ds}{s}\,\Tr\,e^{-sp^2}\,e^{s{\cal A}^2}}\\ 
\nn
&\times & \left\{\frac{s^2}{2}D^2\, D^2 + \frac{(2is)^2\, s^2}{3!}p_i\,p_j\,\left[D^2 D_i D_j 
+ D_iD^2 D_j + D_i D_j D^2\right] 
+ \frac{(2is)^4}{4!}p_i\,p_j\,p_k\,p_l\,D_i\,D_j\,D_k\,D_l \right\}\;.
\eea
For the integration over momentum we use the following relations:
\bea\label{d3pint}
\int\frac{d^3 p}{(2\pi)^3} e^{-s p^2} &=& \frac{1}{(4 \pi s)^{3/2}}\; ,\\ 
\int\frac{d^3 p}{(2\pi)^3}\,p_i\,p_j\, e^{-s p^2} &=&\frac{1}{2 s} \frac{1}{(4 \pi s)^{3/2}}\,\delta_{ij}\;
,\\
\int\frac{d^3 p}{(2\pi)^3}\,p_i\,p_j\,p_k\,p_m\, e^{-s p^2} &=&\frac{1}{(2 s)^2} \frac{1}{(4 \pi
s)^{3/2}}\,\left[\delta_{ij}\delta_{km} + \delta_{ik}\delta_{jm} + \delta_{im}\delta_{jk}\right]\;,
\eea
to obtain
\be
\left[\log\,\det (-D^2)_{\rm{n}}\right]_{{\rm M,\, ghost}}^{(2)}= - \frac{1}{8\,\pi^{3/2}}\int\,d^3
x\,\sum_{k=-\infty}^{\infty}\int_0^{\infty}\frac{ds}{\sqrt{s}}\,\Tr\,e^{s\A^2}\left\{\frac{1}{12}[D_i,D_j]\,[D_i,D_j]\right\}\;.
\ee
As $[D_i, D_j] = - i\,F_{ij}$ the commutator squared gives
\be
[D_i, D_j]^2 = -F_{ij}F_{ij} = -2 B_k B_k\;,
\ee
which leads us to
\be
\label{ghmag}
\left[\log\,\det (-D^2)_{\rm n}\right]_{\rm M, \,ghost}^{(2)}=
\frac{1}{8\,\pi^{3/2}} \frac{1}{12} 2\int d^3x\,
\sum_{k=-\infty}^\infty \int_0^\infty \frac{ds}{\sqrt{s}}\,\Tr\left(e^{s\A^2}\,B_k B_k \right)\,.
\ee
 
For the gluons we get the same result times a factor of $-2$ and one
additional term
\be
T_4\equiv\Tr \int_0^1\,d\alpha\,\int_0^{1-\alpha}\,d\beta\,e^{\alpha s \A^2} (2s\epsilon^{acb}F_{ij}^c)\,e^{\beta s
\A^2}(2s\epsilon^{dfe}F_{ij}^f)\,e^{(1-\alpha-\beta) s \A^2}\;, 
\ee
which with $F_{ij}^a F_{ij}^b = 2 B_k^a B_k^b$ and after integration
over $\alpha$ and $\beta$ yields
\be
T_4 =  4\,s^2\Tr\left(e^{s\A^2} B_k B_k\right)\;.
\ee
Its contribution to the action in the magnetic sector (\ref{Wexp}) 
\be
\frac{1}{2}\int d^3 x\int_0^{\infty}\frac{d^3 p}{(2
\pi)^3}\sum_{k=-\infty}^{\infty}\int_0^{\infty}\frac{ds}{s}\,e^{-sp^2}\,T_4 =
\frac{2}{8\,\pi^{3/2}}\int\,d^3 x\,\sum_{k=-\infty}^{\infty}\frac{ds}{\sqrt{s}}\,\Tr\left(e^{s\A^2}\,B_k
B_k \right)
\ee
is hence of the same structure as the ghost contribution, but multiplied by a
factor of $12$. In the last expression we used eq. (\ref{d3pint}) for the $p$-integration.
Adding the two contributions of the gluon determinant we find that $\left[\log\,(\det
W)^{-1/2}_{{\rm n}}\right]_{{\rm M}}^{(2)} = 10\times \left[\log\,\det (-D^2)_{\rm{n}}\right]_{{\rm
M}}^{(2)}$.

\subsection{Integrating over $\alpha$, $s$, $p$ and summing over Matsubara frequencies}
We have to integrate over $s$ and to sum over the Matsubara frequencies. For
the ghost contribution (\ref{ghmag}) we find that it is apart from the factor
in front of the integral equal to the second invariant, that we computed for
the electric sector, if we replace electric field by magnetic field. For the
precise coefficients, we have to compare eq. (\ref{ghmag}) with the results for
the invariant $I_2$ after the momentum integration (\ref{Q1}) and (\ref{Q3}).
For the ghost determinant in the magnetic sector this yields $-1/(4\pi^2)$ times the functions
$g_{1,2}$ defined in
eqs. (\ref{g1},\ref{g3}). Keeping in mind that the total
action is $11$ times the ghost contribution, we find:
\be
h_1(\nu) = -\frac{11}{4\pi^2}\, g_1(\nu) \qquad\rm{and}\qquad h_3(\nu) =
-\frac{11}{4\pi^2}\, g_3(\nu)\;.
\ee
To obtain the function $h_2$ defined in eq. (\ref{proph}) we denote $h_2=h_3 - h_1$. 

\end{appendix}


\end{document}